# Olivine-Carbonate Mineralogy of the Jezero Crater Region

A. J. Brown[1], C. E. Viviano[2] and T. A. Goudge[3]

[1]Plancius Research, Severna Park, MD 21146. [2]Johns Hopkins Applied Physics Laboratory, MD. [3]Jackson School of Geosciences, The University of Texas at Austin, TX.

Corresponding author: Adrian Brown (adrian.j.brown@nasa.gov)

**Key Points:**

- We identify a correlation between carbonates in the Jezero crater region and the 1 μm band position of associated olivine
- We use the shape and centroid of the 1 μm band to place bounds on the grain size and composition of the olivine
- We examine the thermal inertia of carbonate and olivine bearing regions and find it is not correlated with the 1 μm band centroid position

**Plain Language Summary**

We used Martian orbital data from the CRISM instrument on the spacecraft Mars Reconnaissance Orbiter to assess the minerals on the surface of Mars at the future landing site of the Mars2020 rover. We found correlations between the minerals olivine and carbonate which are of interest to planetary scientists due to often being associated with volcanic activity (olivine) and life (carbonates). We used spectra of these minerals to infer the likely composition and grain size of the olivine unit for the first time, and we document our new approach.

**Abstract**

A well-preserved, ancient delta deposit, in combination with ample exposures of carbonate rich materials, make Jezero Crater in Nili Fossae a compelling astrobiological site. We use Compact Reconnaissance Imaging Spectrometer for Mars (CRISM) observations to characterize the surface mineralogy of the crater and surrounding watershed. Previous studies have documented the occurrence of olivine and carbonates in the Nili Fossae region. We focus on correlations between these two well studied lithologies in the Jezero crater watershed. We map the position and shape of the olivine 1 μm absorption band and find that carbonates are only found in association with olivine which displays a 1 μm band shifted to long wavelengths. We then use THEMIS coverage of Nili Fossae and perform tests to investigate whether the long wavelength shifted olivine signature is correlated with high thermal inertia outcrops. We find no correlation between thermal inertia and the unique olivine signature. We discuss a range of formation scenarios, including the possibility that these olivine and carbonate associations are products of serpentinization reactions on early Mars. These lithologies provide an opportunity for deepening our understanding of early Mars, and, given their antiquity, may provide a framework to study the formation of valley networks, and the thermal history of the martian crust and interior from the early Noachian to today.

# 1 Introduction

### *1.1 Martian Carbonate deposits*

In the Martian context, it is the relative absence of carbonate over most of the globe that makes the mineral so intriguing. Carbonates were predicted to be present on Mars as large sedimentary deposits and have been sought by remote sensing studies as evidence of basaltic weathering under a greenhouse atmosphere in Mars' distant past. Pollack et al. (1987) ran a 1D climate model of Mars under a $CO_2$ rich atmosphere and found the only way for Mars to warm early in its history and retain liquid water on the surface was to somehow form a 1-5 bars $CO_2$ atmosphere and thus create a global greenhouse. They used a terrestrial silicate weathering model to show that this atmosphere would likely disappear through weathering of basaltic material on the surface, creating abundant carbonate-bearing deposits on the surface. Pollack et al. also suggested that to extend the lifetime of the thick atmosphere, the carbonate rocks may have been recycled back into the atmosphere either through burial and thermal decomposition or direct decomposition through contact with hot lava. They stated: "*A test of this theory will be provided by future spectroscopic searches for carbonates in Mars' crust*". Recent studies (Bandfield et al., 2003; Edwards and Ehlmann, 2015) have placed some limits on the amount of carbonate exposed on the surface of Mars as detected by current orbital instruments and small amounts of carbonates have been found in the Comanche rock at Gusev crater (Morris et al., 2010; Carter and Poulet, 2012), however the hypothesized abundant carbonate-bearing deposits remain elusive.

The largest surface exposure of carbonate-bearing material identified to date was discovered in the Nili Fossae region using the CRISM instrument on the MRO spacecraft (Ehlmann et al., 2008b). The origin of these carbonate deposits has been the subject of continuing debate in the Martian scientific community (Niles et al., 2012; Wray et al., 2016). The carbonate is always associated with an olivine spectral signature, as indicated by a wide 1 μm band, and we will therefore use the term "olivine-carbonate lithology" to describe this mineralogical association. The carbonate absorption bands also sometimes occur with phyllosilicate Mg/Fe-OH hydroxyl bands.

The consistent co-occurence of Nili Fossae carbonates and Mg/Fe-phyllosilicates could have many physical explanations. Ehlmann et al. (2009) suggested low grade metamorphic or hydrothermal alteration in neutral to alkaline conditions may have formed these assemblages. Serpentinization is an especially promising method of simultaneous formation of carbonate and Mg/Fe-phyllosilicate (typically serpentine/greenalite or talc/minnesotaite) (Brown et al., 2010; Klein et al., 2013). Viviano et al. (2013) discovered that these assemblages are limited to the eastern Nili Fossae region, and suggested that they may have formed through continued alteration via high temperature (> 200°C) or low-temperature (≤ 200°C), fluid-limited carbonation of serpentine to form magnesium carbonate and talc-bearing material, in contact with the underlying Noachian mixed-layer clay. McSween et al. (2014) pointed out that at the expected alteration temperatures of <350ºC, serpentine is favored for low $CO_2$ activity, whereas talc is stable at higher $CO_2$ activity (> 0.1) conditions in the alteration fluid. Higher $CO_2$ activity tends to produce large amounts of carbonate in the final serpentinization assemblage (Greenwood, 1967; Hemley et al., 1977). Amador et al. (2018) recently conducted a search for serpentinizing minerals using a globally distributed subset of the CRISM dataset and showed these assemblages are restricted to eastern Nili Fossae in association with olivine-rich basalts.

Four related studies have invoked a shallow subsurface, low temperature emplacement of carbonate on Mars using $CO_2$ rich brines. Niles et al. (2005) studied carbonate isotope ratios from ALH84001 and presented two models of $CO_2$ bearing fluids traveling into the surface and create brine like conditions that subsequently form carbonate rocks, invoking fluid mixing and an atmospheric $CO_2$ origin to explain their isotopic observations. Michalski and Niles (2010) presented evidence of layered phyllosilicates and carbonates in nearby Leighton Crater and suggested they may have been buried by Syrtis Major lavas and then excavated from 6 km deep and may represent a regional subsurface carbonate reservoir. Glotch and Rogers (2013) conducted a study of the TES dataset searching for breakdown products of carbonates, specifically creating an index for identification of the Ca-carbonate breakdown product portlandite. Their automated factor analysis (PCA) based search achieved the best matches in the Nili Fossae region. Following the ideas of Michalski and Niles, they proposed a subsurface formation model that was driven by heat from Syrtis Major lavas, resulting in thermal metamorphism and assimilation of the carbonates and eventually production of portlandite. Finally, van Berk and Fu (2011) produced a dynamical geochemical model to study the subsurface layering caused by low temperature reactions under a thick $CO_2$ atmosphere. This model was used by Edwards and Ehlmann (2015) to suggest that that low temperature carbonate sequestration was not likely to have removed sufficient $CO_2$ from an early Mars atmosphere capable of causing a global greenhouse.

*1.2 Nili Fossae and the Jezero Crater paleolake*

This study concentrates on the ancient Nili Fossae region of Mars, which is located adjacent to the Isidis impact basin (Figure 1). The peak of large basin formation in Martian history is estimated at between 4.1-4.25 Ga (Frey, 2008). The Isidis impact basin, at 1350 km across, is the youngest large basin currently recognized and is dated at 3.85-3.8 Ga using observed crater retention age and Hartmann-Neukum model ages (Frey, 2008). The Noachian impact of the Isidis bolide created concentric features in the Martian upper crust that were discovered by Mariner 9 (McCauley et al., 1972) and soon thereafter were confirmed to be crater-related features (Wilhelms, 1973). The Isidis multi-ring impact structures are represented today as massive troughs that are exposed to the northwest of the Isidis Planitia basin (Schultz and Frey, 1990). These features have been modeled as relaxation graben structures caused by the weight of the Isidis mascon (Comer et al., 1985; Ritzer and Hauck, 2009). The features are known as the Nili Fossae, and they give their name to this well exposed, largely dust free, chunk of Noachian crust in which they lie. The low relief, early Hesperian shield volcano of Syrtis Major lies to the south, and its lavas onlap the southern Noachian exposures of the Nili Fossae and the rim of the Isidis Basin (Hiesinger and Head, 2004).

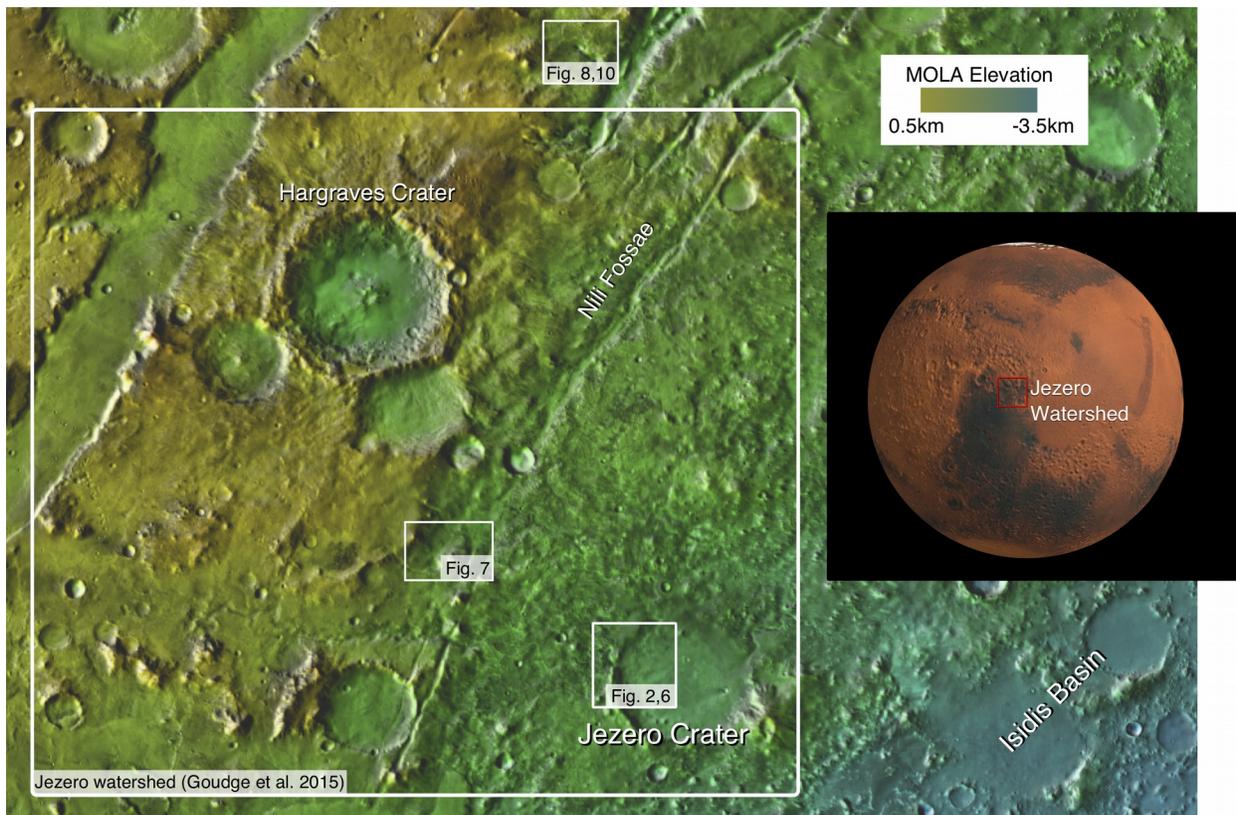

**Figure 1.** Location of Nili Fossae and Jezero crater showing regions discussed in the text. Base map is "THEMIS Day IR with MOLA Color" obtained using JMars (Christensen et al., 2009). Inset image shows the study region in a red box, with the location shown as relatively dark and dust free.

Amongst the sites of interest in the Nili Fossae region is the paleolake basin contained within the ~45 km diameter Jezero impact crater (Fassett and Head, 2005). The Jezero crater paleolake is classified as hydrologically open, and was fed by two inlet valleys to the north and west, and drained by an outlet valley to the east (Fassett and Head, 2005). Buffered crater counts of inflowing valley networks indicate that this system ceased fluvial activity by approximately the Noachian-Hesperian boundary (Fassett and Head, 2008), similar to the timing of other large valley network systems on Mars (Fassett and Head, 2008; Hoke and Hynek, 2009). Jezero crater contains two well-exposed fluvio-lacustrine delta deposits (Ehlmann et al., 2008a; Fassett and Head, 2005; Goudge et al., 2017, 2015; Schon et al., 2012) as well as large exposures of both phyllosilicate minerals and carbonates (Ehlmann et al., 2009, 2008b, 2008a; Goudge et al., 2017, 2015). *In situ* exploration of this region could provide novel insights into both the fluvial sedimentary record and aqueous alteration history of early Mars. These outcrops contributed to the selection of Jezero crater in November 2018 as the landing site for the Mars 2020 rover (Williford, 2018).

An idealised geological sequence of events at Jezero crater is shown in Figure 2. This Figure outlines our current interpretation of the timing of major geological processes that have shaped the crater and surrounding watershed as summarised in four steps. Step 1 is the impact that formed Jezero Crater, which postdates the Isidis basin formation and formation of megabreccia in the Nili Fossae regional basement unit (Mustard et al., 2009). Step 2 is emplacement of an olivine bearing lithology and partial carbonatization of this lithology. Step 3 is the formation of fluvial valleys, filling of the crater with water and the emplacement of deltaic deposits, during the Late Noachian-Early Hesperian valley network forming period (Fassett and Head, 2008; Irwin et al., 2005). Step 4 is the erosion of the delta deposits to their current degraded state and infill of the crater by the floor resurfacing unit (Goudge et al., 2015; Schon et al., 2012). The latter two steps must post-date the carbonate-formation, as the northern and western delta appear to contain detrital carbonate transported to the basin from the watershed (Goudge et al., 2015).

The two delta deposits within Jezero contain Fe/Mg-phyllosilicate and carbonate in varying proportions, with the northern fan dominated by carbonate and the western fan dominated by phyllosilicate (Goudge et al., 2015). The provenance of the phyllosilicate and carbonate within the Jezero crater deltas can be traced to mineralogically similar protolith units within the watershed, which provides strong evidence that the alteration minerals were primarily emplaced by fluvial transport (Goudge et al., 2015). These deposits therefore have bulk compositions that integrate heavily altered martian crust of the Nili Fossae region (Ehlmann et al., 2009; Goudge et al., 2015; Mustard et al., 2009), and offer an opportunity to examine a diverse array of alteration minerals. Furthermore, Jezero crater contains large exposures of olivine- and carbonate-bearing units on the floor of the crater, underlying the deltaic deposits, and draping the interior rim of the crater (Ehlmann et al., 2009, 2008b; Goudge et al., 2015). These deposits were interpreted by Goudge et al. (2015) to represent exposures of the regional olivine-carbonate-bearing unit observed elsewhere in Nili Fossae (Ehlmann et al., 2009, 2008b) and are similar to carbonate-bearing geomorphic units mapped recently in a study of the neighboring northeast Syrtis Major region (Bramble et al., 2017). A recent study by Salvatore et al. (2018) used regional thermal infrared observations to derive modal mineralogy of the N.E. Syrtis and Jezero crater regions and derived average carbonate abundance (4-17%, their Figure 11) for 12 TES spectra collected over Nili Fossae.

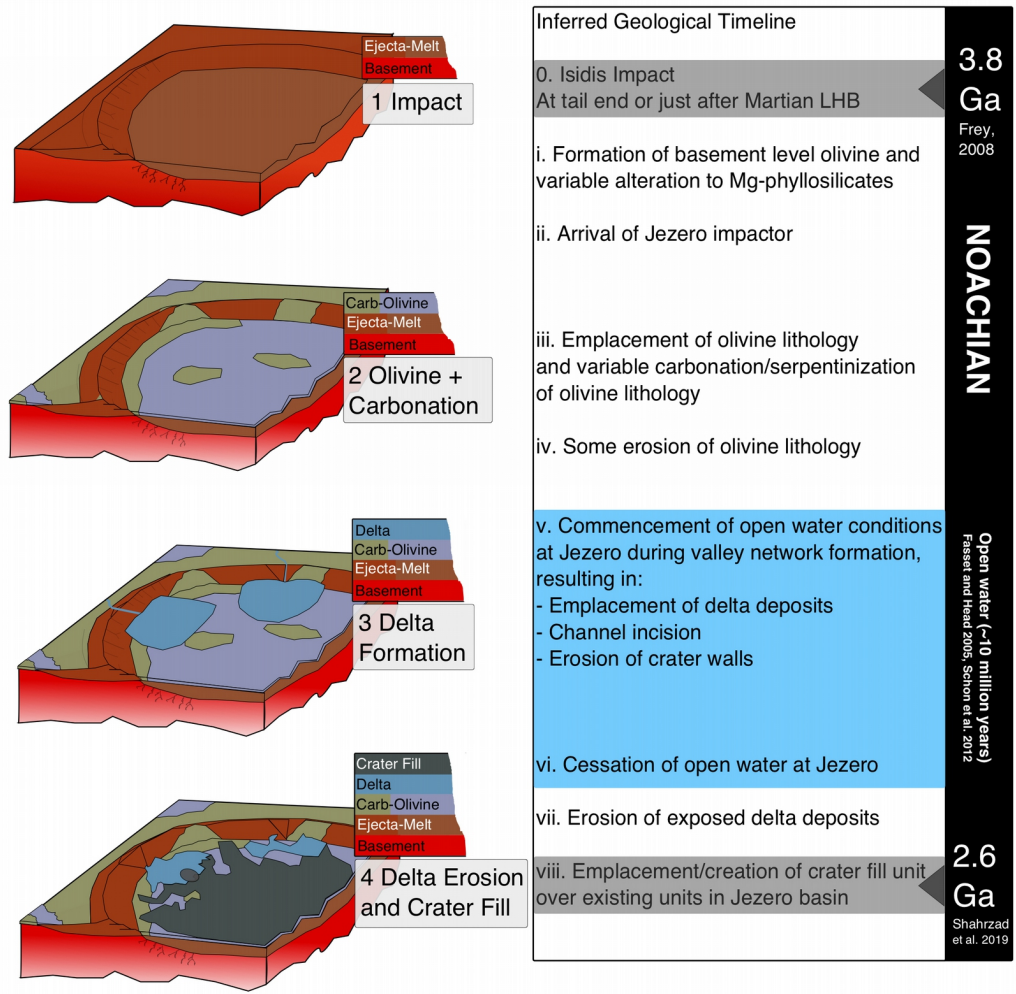

**Figure 2.** Framework formation history for Jezero crater lithologies for the period 3.8-3.2 Ga. Step 1 is the impact that formed Jezero, Step 2 is the emplacement of the Fe-rich olivine-carbonate lithology and the variable carbonatization of this lithology, Step 3 is the emplacement of the two deltas and Step 4 is erosion of the deltas and emplacement of crater infill, leading to the crater as we see it today. Age of Isidis impactor is from Frey (2008) and age of crater fill unit is constrained by crater study of Shahrzad et al. (2019).

### *1.3 Previous Studies of Martian Olivine*

Olivine, [(Mg,Fe)$_2$SiO$_4$], is an important primitive planetary mineral which exhibits a continuous spectrum between its Fe-rich endmember Fayalite (Fa) and Mg-rich endmember Forsterite (Fo). Olivine-dominated (>20%) lavas (picrites) have been detected *in situ* by MER *Spirit* at Gusev crater (McSween et al., 2006). McSween et al. reported the APXS estimates of the composition of the olivine as measured on basaltic "subcrop" rocks Adirondack, Humphrey, and Mazatzal as Fo52, 49, and 45 respectively (see Figure 3). They used the pMELTS program to predict the original mantle olivine composition range of these three rocks and found it possible that these basalts formed in a part of the mantle with Fo55-81 composition. The MER team also reported Mossbauer measurements that estimated the olivine composition of these olivine rich rocks as Fo60 (Morris et al., 2004). Finally, using the Mini-TES instrument, Christensen et al. (2004) provided an estimate of Fo35-60. These MER instrument olivine compositional findings are shown in Figure 3.

McSween et al. (2006) compared the findings of the MER team to the known meteorites and determined that the most similar family are the olivine-phyric shergotitites. They pointed out that the shergotitites are too young to be linked to the ancient Columbia Hills basalts, however they are the only meteorites family with 1) similar olivine phenocrysts, 2) similar olivine composition - (Gusev: Fo40-60, olivine-phyric shergotites: Fo25-84 with norm of 65), 3) similar modal abundances (both 20-30 vol%) and 4) similar coexisting mineralogies. The olivine-phyric shergotitites compositional ranges are also summarized in Figure 3.

The composition of the Martian average mantle and crust (excluding the core) has been estimated to be ~Fo77 from SNC meteorite compositional abundances (Dreibus and Wänke, 1985) compared to the terrestrial mantle which is estimated to be ~Fo89 (Wänke, 1981). This implies that an undifferentiated lava direct from the Martian mantle would have a more Fe-rich composition compared to their terrestrial counterparts. In order for higher Mg mantle derived lavas to erupt, hotter conditions must be invoked, including hot spot volcanism. Filiberto and Dasgupta (2015) calculated that an increase in 1 Fo number corresponds to addition of 20 K to the temperature of the mantle source. Cooler magma source conditions or mixing with wallrock or eruptions sourced from thin crustal regions (such as due to the Isidis impact) are hypothesized to play a role in controlling the temperature of the mantle source for Nili Fossae basalts (Kiefer, 2005).

A magma ocean model that places constraints on the olivine composition of the early Martian crust was proposed by Elkins-Tanton et al. (2005). This study presented a deep magma ocean model that predicted variations in the ancient Martian crust to 2000 km depth. As envisaged by the model, in the 30-50 Mya after the accretion phase of Mars, a solid state, gravitationally driven overturn, would take place to stabilize the Martian interior. The model suggested that mid-mantle olivine would melt and result in a Mg-rich (Fo87-89) layer at the top 200km of the Martian crust. The model predicted a mid-mantle layer of Fo60-78 at 1000 km depth, and an inaccessible Fe-rich (Fo25-59) 1500-2000 km deep beneath the surface. Shortly after the Elkins-Tanton model was proposed, Grott and Breuer (2008) showed that high heat fluxes of the early Noachian would lead to small elastic lithosphere thickness and thin boundary layers in the upper mantle, consistent with mantle convection that was active and vigorous at this time. As a result the stable mantle stratification of the magma ocean could not have lasted more than 300-500 Myr. Baratoux et al. (2013) also argued that the early Martian mantle would have been hotter and more effusive than envisaged in the Elkins-Tanton model, and used mineralogical models of observed lithologies at Noachian sites to suggest that voluminous Noachian lava fields have subsequently covered up evidence of the posited magma ocean. We will discuss these models further as we proceed.

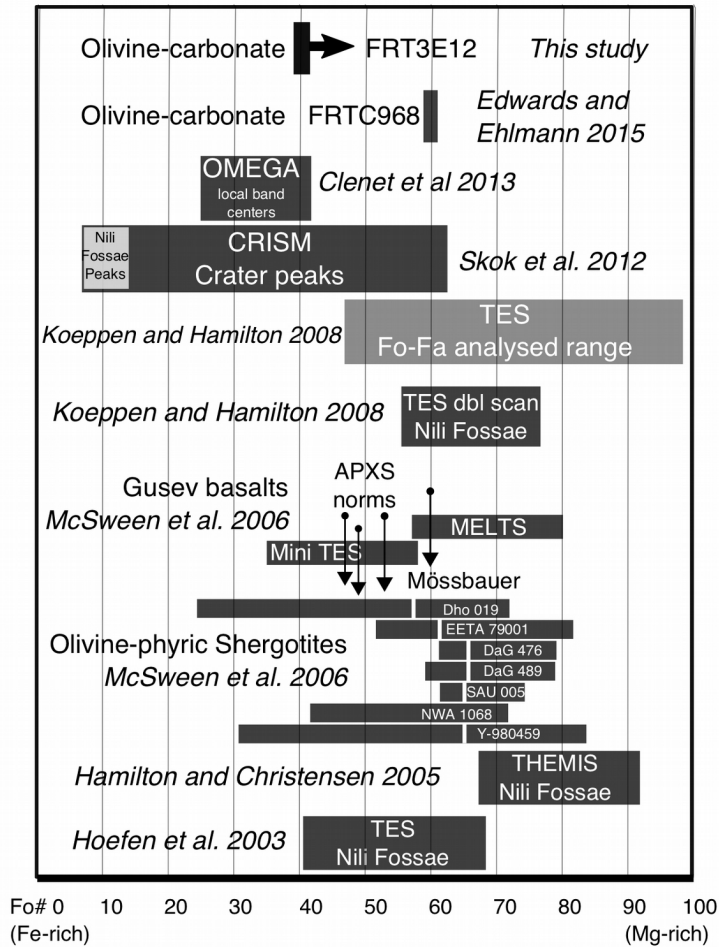

**Figure 3.** Summary of olivine composition results from previous remote and *in situ* studies of global Mars surveys, MER measurements at Gusev crater and the olivine-phyric shergotites, updated and expanded from McSween et al. (2006).

Fayalite is 10 times less stable than Forsterite to low temperature weathering, which is thought to be relevant to modern Mars (Stopar et al., 2006). Fa is more dense than Fo, which may lead to enrichments of fayalite in lag deposits under density (hydrodynamic) sorting due to aeolian processes (Fedo et al., 2015). Spectroscopically, Fa is discriminable from Fo because the substitution of $Mg^{2+}$ for $Fe^{2+}$ in the M1 and M2 sites of olivine crystal structure changes the site configuration and distorts the shape and position of the absorption band in energy space (Burns, 1970; Sunshine and Pieters, 1998; King and Ridley, 1987; Isaacson et al., 2011). The occupancies of the M1 and M2 sites both effect the 1 μm band that we use in this study, therefore the band distortion is a not straightforward shift to long or short wavelengths. In addition, $Mn^{2+}$ substitution can take place in these locations (Burns and Huggins, 1972), however we ignore that process for this study as Mn because although it is concentrated in some rocks at Gale crater (Lanza et al., 2014), it is a minor component (~0.4 wt % MnO) compared to Mg and Fe in most Martian basalts (Taylor and McLennan, 2009). As will be discussed below, the strongest factor in shifting the 1 μm band to shorter wavelengths is high $Mg^{2+}$ content, the strongest factor pushing it to longer wavelengths is higher $Fe^{2+}$ content.

Olivine was directly discovered on Mars in the Nili Fossae region using the Thermal Emission Spectrometer (TES) instrument by Hoefen et al. (2003). Hoefen et al. used a spectral comparison approach to identify the olivine 400 cm$^{-1}$ wavenumbers (25 µm) band in several TES spectra over Nili Fossae (at 3x5 km/pixel resolution). They showed that the 25 µm band shifts with Fo#, shifting to longer wavelengths for higher Fe content. They compared the band positions to those in the King and Ridley (1987) spectroscopic study of olivine composition. They reported best matches of the spectra to Fo66, with some at Fo41 and some Fo60. Hoefen et al. looked at the 25 µm band for four GDS70 olivines (Fo89 from Green Sand Beach, Hawai'i) which were sieved and separated into grain size distributions of <60, 60-104, 104-150 and 150-250 microns. No significant shift in 25 µm band position was observed for these grain size separates. They therefore compared all their spectra to three Kiglapait (KI Fo66-41) samples which were <60 microns grain size. Hoefen et al. found the best matches to Fo66 in the southwest and down to Fo41 in the northeast of Nili Fossae. They did not map Low-iron olivine in the Nili Fossae region.

In the wake of the olivine discovery at Nili Fossae, Hamilton and Christensen (2005) conducted a study using a decorrelation stretch of THEMIS images to map olivine bedrock across a large swath of Nili Fossae. They carried out a difference technique to constrain the composition to be Fo68-91, and indicated that their compositional range was more magnesian and broad, due to the lower spectral resolution of THEMIS. They also identified layering and digitate forms associated with this olivine unit and used this observation to infer a magmatic origin.

Koeppen and Hamilton (2008) then used the TES dataset to extend the original findings of the Hoefen study to a global map. They developed an index to find the signature of five Fo# ranges in a filtered subset of the TES dataset. Koeppen and Hamilton only used one grain size range, (710-1000 microns), which they stated was appropriate to dark regions such as Nili Fossae according to the model results of Ruff and Christensen (2002) which suggested material this dark material was consistent with larger grain size material. Koeppen and Hamilton pointed out that for low Fo# olivines, the TES spectra are sensitive to atmospheric interference and also instrumental noise. As a result, they only presented index maps for the three Fo# groups: Fo75-100, 58-74 and 42-57. Koeppen and Hamilton also performed a 500 km$^2$ "double scan average" to reduce instrumental noise over Nili Fossae and obtained a qualitative match for their Fo58-74 index. They produced a filtered global map of Mars that includes coverage of Nili Fossae. These maps show some Fo75-100 detections in and around Nili Fossae, however the authors stated that most of the spectra from the region appear to best match Fo58-74. Koppen and Hamilton concluded that their finding of relatively high Fe olivine called into question the Elkins-Tanton et al. (2005) magma ocean model, which envisions an early crust with Mg-olivine lying above Fe-olivine. They put forward an alternate model that includes an extra "leftover magma" that would supply an Fe-rich liquid and form a secondary crust above the uppermost Mg-olivine layer from the magma ocean. They pointed out that this is in better accord with their observations of relatively more Fe-olivine in the TES global dataset.

Previous studies using Visible and Near Infrared (0.35-5.1 µm) data collected by the Observatoire pour la Minèralogie, l'Eau, les Glaces et l'Activitè (OMEGA) instrument on Mars Express have determined the presence of olivine in the Nili Fossae region. Mustard et al. (2005) used OMEGA data to map olivine and pyroxene in the Nili Fossae region, noting variations in the apparent position of the olivine 1 µm band position, possibly due to variations in grain size.

In a follow up study using higher resolution CRISM data, Mustard et al. (2009) showed a fayalite-like spectrum from CRISM image FRT00003E12 in their Figure 3, and noted that large particle size and textural effects could create the observed broad absorptions.

A number of global surveys using the OMEGA dataset that have mapped olivine composition have also been completed. Poulet et al. (2007) developed a set of spectral parameters for the OMEGA global dataset and mapped an Fo and Fa-endmember in a global map for the first time. They again emphasized that large grain size may shift the olivine 1 μm band to the longer wavelengths. They reported that the Fa-endmember, while easy to detect, was generally limited in occurrence and areal coverage. At the global scale, it was predominantly detected in the Nili Fossae region (Poulet et al., 2007).

In a follow on to Poulet's et al. study, Ody et al. (2013) conducted a further global study using OMEGA spectral indicies and divided their olivine detections into Type 1 and 2. Their Figure 4 shows Type 1 olivine, which they described as small grain size or Mg rich, was far more common globally, and did occur, to a limited extent, in the Nili Fossae region. Type 2 olivine, which they described as Fe rich and/or large grain sized, was easier to detect, but was far more limited in extent. Their Figure 4c also shows the greatest concentrations of Type 2 olivine are at Nili Fossae.

Clenet et al. (2013) conducted a survey of OMEGA data using the Modified Gaussian Method (MGM) to constrain the mafic mineralogy. They reported that locally, within Nili Fossae, Fe-rich olivine was present, however their study used lower spatial resolution OMEGA data and they were not able to detect or correlate the presence of carbonate with olivine. Clenet et al. (2013) found that, locally, band centers are shifted to longer wavelength which could indicate Fo25-40 compositions or larger grain size.

Skok et al. (2012) used the MGM approach to calculate the Fo# of CRISM observations of Martian crater central peaks, in order to access the most primitive Martian crust samples, and found a range of Fo5-60 in the craters they studied. In the Nili Fossae region (at Hargraves crater and an unnamed crater), they found extremely iron-rich olivine compositions of Fo5-14. Skok et al. (2012) noted that Clenet et al. (2011) had found a spectral dependence of grain size but did not address the possibility of grain size affecting the position of the olivine 1 μm band and the possible effect on their MGM results. The olivine Fo# results we have discussed are summarised (where compositional estimates were provided) in Figure 3.

### 1.4 Outline of the paper

The overarching goal of this study is to use CRISM data to map at high spatial and spectral resolution occurrences of previously detected carbonate and olivine deposits in order to determine whether any correlation exists at the 20 m scale between these two well-studied lithologies present in the Nili Fossae region. We then assess whether any relationship exists between thermal inertia (using THEMIS data) and the olivine-carbonates detected by CRISM. We then summarize the potential effects of various physical factors on the 1 μm band position and draw conclusions from this. Finally, we study a variety of carbonate emplacement methods that have been proposed in the literature and determine which are more consonant with the mineralogical associations we have discovered.

## 2 Methods

We have analyzed data from the CRISM instrument (Murchie et al., 2007) covering the Jezero crater and the associated watershed for the presence of carbonate and olivine minerals exposed at the surface. A detailed list of CRISM observations and mineral identifications is listed in the Supplementary Material accompanying this paper. Our approach to mapping olivine composition largely follows the example of previous studies of the olivine composition of the Moon using continuum removed M$^3$ data (Isaacson et al., 2011) and follow-on laboratory studies of synthetic olivine samples (Isaacson et al., 2014) using the MGM approach (Sunshine and Pieters, 1998). These studies have established the ability of continuum removed VNIR spectra to determine the relative Fe vs. Mg composition from orbit.

CRISM has two primary operational modes, gimbaled hyperspectral (targeted: FRT, HRS, HRL) and push broom multispectral (mapping: MSP, HSP, MSW), that allow for both targeted hyperspectral coverage over large areas of interest, and global multispectral coverage at reduced spatial resolution. The multispectral mapping data are derived utilizing 72 channels of the hyperspectral set that have been carefully chosen to heavily sample wavelength regions in the infrared where narrow bands associated with alteration minerals occur. We use prototype or publicly available Map-projected Targeted Reduced Data Record (MTRDR) CRISM products for all full and half resolution images. These products have gone through a series of spectral corrections and spatial transformations to remove atmospheric absorptions and systematic noise, and normalize effects of atmospheric dust (Seelos et al., 2016).

### 2.1 Olivine 1 µm band and Asymmetric band fitting

In this study we have primarily used the Asymmetric Gaussian spectral band fitting approach (Adrian J. Brown, 2006; Brown et al., 2010) to produce olivine discriminating maps of the Jezero crater watershed. We summarize the key points of the model here, discuss the parameters of this model and provide example Asymmetric Gaussian fits. Further discussion of the parameters for each step is provided in the Supplementary Material.

The Asymmetric Gaussian model achieves an iterative best fit using a Nelder-Mead (1965) simplex routine to optimally update the parameter set. The function to be fitted is an Asymmetric Gaussian shape:

$$\text{If } \lambda \leq \lambda_0 \quad f(\lambda) = \alpha \exp\left(-\left[\frac{\lambda - \lambda_0}{\sigma^2}\right]^2\right) \qquad \text{if } \lambda > \lambda_o \quad f(\lambda) = \alpha \exp\left(-\left[\frac{\lambda - \lambda_0}{(\chi \sigma)^2}\right]^2\right) \qquad (1)$$

The parameter set includes the centroid $\lambda_0$, amplitude $\alpha$, half width half maximum (HWHM) $\sigma$, and the asymmetry parameter $\chi$. The asymmetry parameter is unbounded – values less than 1 indicate right asymmetry, values greater than 1 indicate left asymmetry. All four of these parameters contain potentially useful information and can be used to discriminate bands of variable shape and strength from noise. We used the Asymmetric Gaussian approach to model the 1 µm band of two laboratory spectra, selecting the shoulders to be fixed at 0.7 and 1.75 µm. We fit a straight line continuum between these two shoulder points, in order to reduce the number of fitting variables. A step by step plot of the process of fitting three olivine spectra (two laboratory spectra and a CRISM spectrum) with their Asymmetric Gaussian fits is presented in Figure 4.

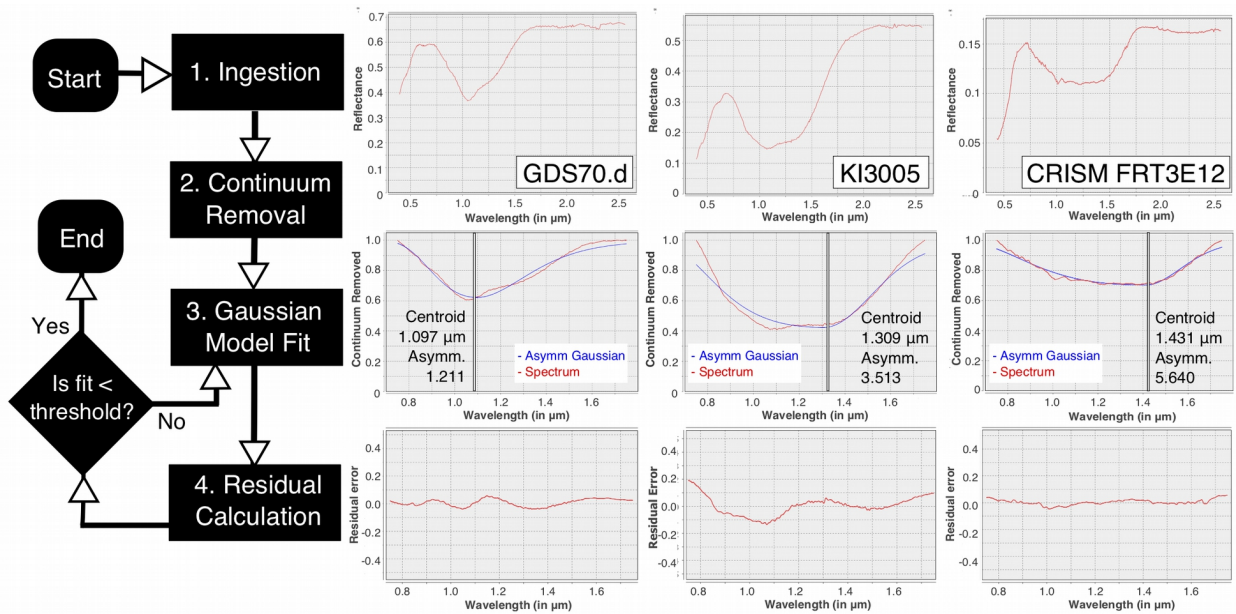

**Figure 4.** Example Asymmetric Gaussian modeling of the library spectra of two laboratory olivine samples (GDS70.d and KI3005) and a long wavelength shifted olivine CRISM spectrum from FRT3E12. The GDS70.d has Fo#89 and the fitting band is more symmetric. The KI3005 has Fo#11 and the fitting band is more asymmetric. The CRISM spectrum 1 μm band is more saturated than the lab spectra and so the centroid and asymmetry are even higher than the lab spectra.

The position of the 1 μm olivine band is sensitive to the Mg vs. Fe content, where more Fe-rich olivine displays a longer (right-shifted) 1 μm band position (King and Ridley, 1987). In order to use the Asymmetric Gaussian model to map the olivine 1 μm band (which is made up of three overlapping Gaussian bands), we have to be judicious in our use of fit parameters and insert reality checks into the fitting process, while at the same time attempting to remove bias and consistently fit CRISM scenes that were taken under variable dust conditions and have some remaining amount of residual noise. We will discuss the external effects of variation of grain size and mixing with other components on this estimate in the Discussion section below.

We obtained from the USGS Spectral Library (Clark et al., 2007) seven Kiglapait (KI – Fo11-66) and four Green Sand Beach (GSB - Fo89) olivine library spectra of varying composition and grain size from the study of King and Ridley (1987). In addition, we supplemented this collection of 11 USGS olivines with 2 additional USGS provided olivine spectra and 8 additional olivines sourced from the RELAB reflectance database (http://www.planetary.brown.edu/relab/), in order to avoid biases due to reflectance measurements in a particular lab, as suggested by Trang et al. (2013).

Due to the dependence of the band shape on grain size, we have produced a plot of the 16 of the 21 olivine samples which have <70 micron grain size. Their measured centroid and asymmetry are shown in Figure 5. These plots show that for both RELAB and USGS laboratory olivine spectra, more Mg-rich samples have lower centroid and asymmetry, whereas more Fe-rich laboratory spectra have higher centroid and asymmetry values.

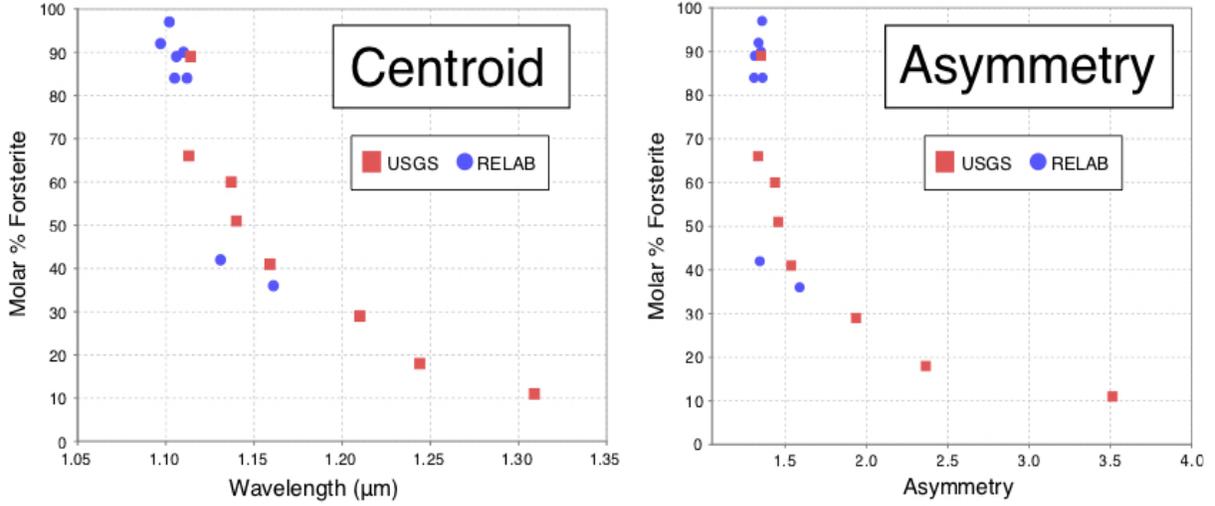

**Figure 5.** Laboratory spectra Asymmetric Gaussian fitting results for RELAB and USGS spectra. These demonstrate the tendency of the centroid and asymmetry to decrease with increasing Fo#. This plot includes all 16 laboratory olivine spectra that have grain size <70 microns.

**Comparison of laboratory spectra to CRISM spectra**. The right column of Figure 4 shows an example CRISM olivine spectrum that has a 1 μm band that is saturated, which causes the band shape to flatten between 1-1.5 μm. For this spectrum, the Asymmetric Gaussian method returns an asymmetry and centroid that are very high. In the middle laboratory spectrum (Fo# 11, ~70 micron grain size), the 1 μm band is deeper than the CRISM spectrum, however it is not as saturated, and this is reflected in the lower centroid and asymmetry than the CRISM spectrum. The occurrence of the saturation and the manner in which it controls CRISM and laboratory spectra is the reason the Asymmetric Gaussian method identifies olivines with high saturation, which are of key interest in this study.

**Rationale of threshold and subsetting approach.** The four parameters derived in the Asymmetrical Gaussian approach can all be utilized to give a more robust estimate for the presence of olivine and its composition. For example, to reduce the chance of noisy spectra adversely affecting the results, minima can be placed on the height of the band, limits can be put on the asymmetry of the band. The choices we make reflect the noise structure and characteristics of the CRISM dataset and are the result of extensive iterative testing for the best parameters across the wide number of CRISM high resolution and mapping observations available to us at Nili Fossae. In order to restrict our analysis to the long wavelength shifted olivine lithology, and to reduce the probability of analyzing noisy spectra, we have thresholded the parameters of the resultant dataset as follows:

$$\text{if } 0.1 < \alpha < 1.0 \text{ and } 1.15 < \lambda_0 < 1.5 \text{ and } \sigma < 0.3 \text{ and } 0.8 < \chi < 9 \qquad (1)$$

where $\alpha$ is the amplitude, $\lambda_0$ is the centroid, $\sigma$ is the half width half maximum (HWHM), and $\chi$ is the asymmetry parameter. These parameters were chosen in a step-by-step variational approach using targeted and mapping CRISM data across Nili Fossae and were found to give the most uniform and robust results in mapping the olivine-carbonate lithology. This parameter filter has been applied to each CRISM dataset we discuss below in order to eliminate noisy and non-olivine-bearing pixels. By restricting the lowest admissible centroid to 1.15 µm, we have eliminated highly Mg-rich olivine (Figure 5) or fine grained olivine spectra. This is done intentionally to focus the study on long wavelength shifted and saturated olivine signatures.

# 3 Results

Three regions with full resolution CRISM coverage from the Jezero crater, Jezero watershed and just north of the Jezero watershed are presented below. The continuum removed example spectra, Asymmetric Gaussian maps of olivine 1 µm position, carbonate maps, and correlation maps between 1 µm band position and asymmetry, overlain with the presence of carbonate are provided for each region. Finally, thermal inertia maps for these regions are presented for comparison with our CRISM maps.

### *3.1 Olivine-Carbonate correlations in Jezero Crater*

Figure 6a shows continuum removed spectra from a CRISM image over the western Jezero delta at three representative points within the crater (locations shown by arrows in the figure). The continuum removed spectra show a distinct shift in the wide olivine 1 µm band. The black spectrum represents the olivine unit that displays the most long wavelength shifted 1 µm band. The blue spectrum represents an example of an olivine-carbonate spectrum that is exposed on the west of the crater, and also displays carbonate bands at 2.3 and 2.5 µm. The red spectrum corresponds to a lithology that contains the most short wavelength shifted 1 µm olivine band, and is sourced from the western rim of the crater.

Figure 6b shows a plot of the asymmetry versus the centroid for the olivine 1 µm band, both fit by the Asymmetric Gaussian curve method for the HRL40FF CRISM image data shown in Figure 6c. The pixels in Figure 6b are color coded for the strength of the 2.5 µm band which is indicative of carbonates (red pixels show the strongest 2.5 µm band). This scatter plot clearly shows there is spatial correlation between carbonates and high 1 µm band centroid in the Jezero crater scene. Figure 6d shows the carbonate browse (CAR) image which highlights carbonate in green, Mg/Fe-phyllosilicates in magenta and other hydrated hydrated minerals in blue (Viviano-Beck et al., 2014). This map demonstrates that the largest carbonate deposits in the crater are associated with the unit that is mapped in the thresholded 1 µm band centroid map in Figure 6c, as is evident in the correlation map of Figure 6b.

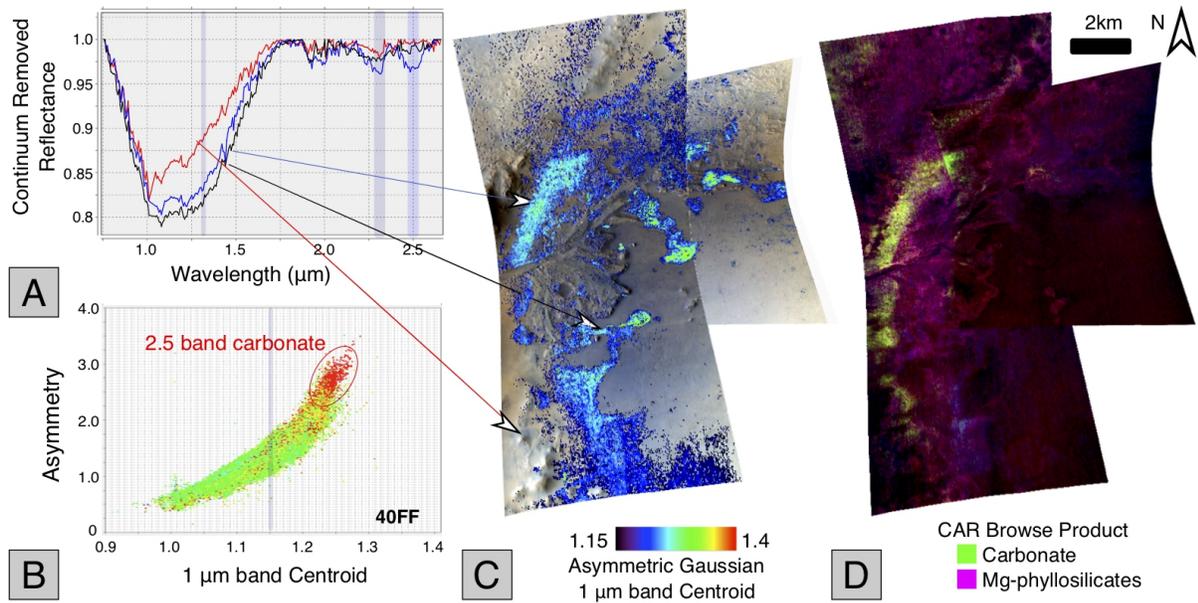

**Figure 6. a.)** Example continuum removed spectra from arrowed locations in images HRL40FF and FRT47A3. The 1.3, 2.3 and 2.5 µm regions are indicated by shaded vertical regions. The blue spectrum shows 2.3 and 2.5 µm bands indicative of carbonates, and the 1 µm band is shifted to longer wavelengths, relative to the red spectrum. **b.)** Plot of asymmetry versus centroid position for the 1 µm absorption band, color coded for the strengths of the 2.5 µm feature indicative of carbonates for all pixels from HRL40FF. The 1.15 µm cutoff for the threshold used in map c) is indicated by a vertical line. **c.)** Olivine 1 µm band centroid map of Jezero western delta covered by CRISM images HRL40FF and FRT47A3. Arrows show regions where example spectra were obtained. **d.)** CRISM CAR standard browse produces of the Jezero delta, showing regions where carbonate is present due to the presence of a 2.3 accompanied by a 2.5 µm band.

### *3.2 Partial Carbonization of olivine-carbonate lithology across Nili Fossae*

Figure 7 presents a CRISM image of a region over the Nili Fossae at the extreme western edge of the Jezero watershed (the location of the image is shown in Figure 1). As for Figure 6, the CRISM 1 µm centroid map has been thresholded to a lower limit to isolate our analysis to only long wavelength shifted olivine-rich exposures.

This figure demonstrates a key property of the carbonation process vis-à-vis the long wavelength shifted olivine-bearing lithology: the carbonate-bearing regions of this lithology are spatially variable. The green arrows on the figure indicate that when the carbonate is present, it is located within a long wavelength-shifted olivine lithology. The white arrows on the Figure 7c and d show regions of the long wavelength shifted olivine-bearing lithology that have not been carbonatized (since this region is not green on Figure 7d).

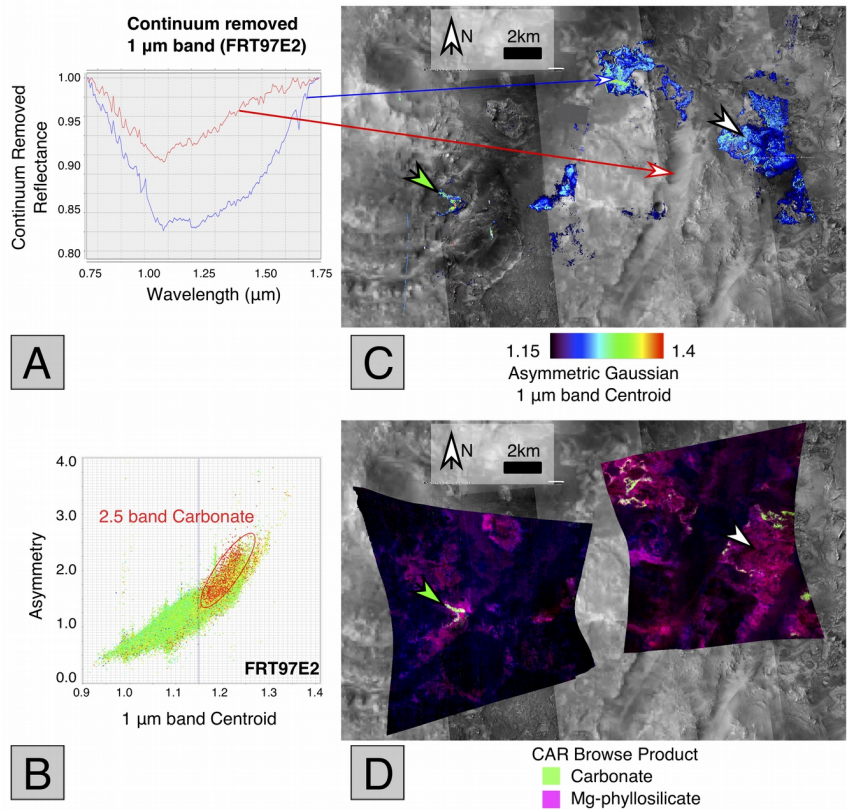

**Figure 7. a.)** Example continuum removed 1 μm band centroid map from two example points showing olivine (blue) and dust with olivine (red). The arrows indicate their locations on the CRISM image. **b.)** The correlation map of carbonate vs. asymmetry and centroid, showing carbonate pixels (with a 2.5 μm band present) in red. The 1.15 μm threshold is indicated by a vertical line. **c.)** Olivine mineralogy of Nili Fossae as mapped by CRISM images FRT23370 (left) and 97E2 (right) overlying a CTX basemap with HiRISE images where available. The 1 μm band centroid map has been thresholded as discussed in the text. **d.)** Carbonate standard browse product (CAR) showing the region where carbonate is mapped. Carbonate is in green and 2.3 μm band material (Mg-phyllosilicate) is in magenta. Note that carbonates are only associated with the longer wavelength shifted olivine unit, and this lithology is only partially carbonatized (green arrow indicates carbonate is detected, white arrows where it is not).

The correlation plot in Figure 7b shows the location of carbonates in the asymmetry vs. centroid space of the olivine 1 μm band for FRT97E2. This provides further evidence that for this image, the carbonates are grouped in a relatively constrained region of the plot, showing above average centroid and high asymmetry values. Two further points are worth noting here:

1.) in this scene, not all pixels that fit this description are carbonate bearing. This spatial relationship is also manifested within Jezero Crater, where the long wavelength shifted olivine lithology is variably carbonatized (see correlation plot in Figure 6b).

2.) in this scene, in contrast to the last, the carbonates do not correspond to the absolute extrema of centroid values in the image, which indicates variation in the olivine-carbonate spatial relationship between watershed and the delta region.

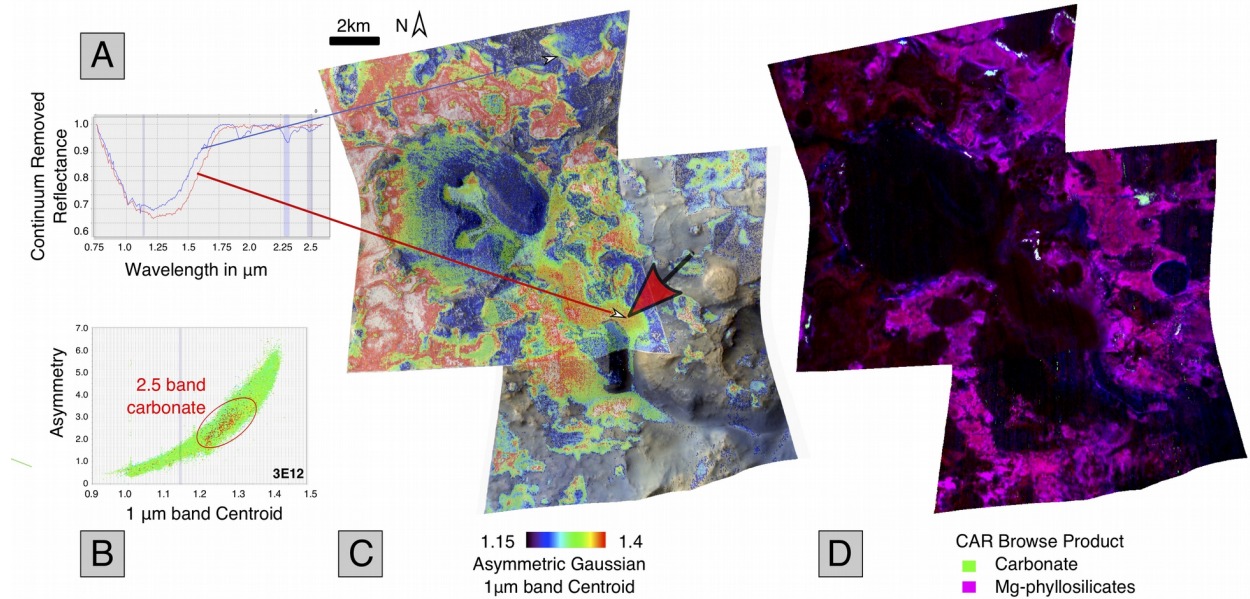

**Figure 8.** Olivine and carbonate mineralogy for CRISM image FRT3E12 and FRTB438. **a.)** Example continuum removed spectra from CRISM FRT3E12, with locations shown by arrows. Note 2.3 and 2.5 bands indicating the presence of carbonate in the blue spectrum. The red spectrum does not have carbonates, and the 1 μm band is further shifted to long wavelengths. **b.)** Asymmetry vs. centroid correlation map for 3E12, color coded in red for the presence of the 2.5 μm carbonate band. The plot shows carbonates are not associated with the extreme right shifted olivine spectra in this scene. The 1.15 μm threshold is shown as a vertical line. **c.)** 1 μm band centroid map for FRT3E12 and FRTB438. **d.)** CRISM CAR Browse product showing carbonates in green and Mg-phyllosilicates in purple.

### 3.3 Evidence for Olivine Band Saturation

Figure 8 presents data for CRISM full resolution images 3E12 and B438, which is located at a re-entrant close to the Nili Fossae trough (see Figure 1 for the location). Figure 8a shows example continuum removed spectra from the scene, including an example with 2.3 and 2.5 μm bands, and one without. Both show long wavelength shifts of the 1 μm band, however the one without carbonate bands is shifted further to long wavelengths. Figure 8a shows an example continuum removed spectrum in red that shows a shallowing of the base of the band that we interpret as band saturation, caused by large grain size, and the same spectrum was analyzed in Figure 4.

Figure 8b is a plot of asymmetry versus centroid position for the 1 μm band, with colors and an ellipse indicating where carbonate bearing material plots. This demonstrates that in this scene, in contrast to the delta scene, but in family with the watershed scene, the carbonate is not associated with extremal shifts in the 1 μm band. Figure 8c shows the 1 μm band centroid images which show that olivine is present in great abundance in these images. Figure 8d shows that the image contains carbonates, mostly outcropping around the central horseshoe shaped feature (which is dark in Figure 8d), exposed on the exterior of this extended mesa.

*3.4 THEMIS Thermal Inertia mapping*

In order to develop a greater understanding of the lithologies we have chosen to focus on in this study, we will now examine the thermal inertia maps generated using THEMIS covering the same regions. To first order, for unconsolidated material, higher thermal inertia corresponds to larger grain size (Fergason et al., 2006), and therefore we wish to establish whether there might be a relationship between the olivine 1 μm centroid position and the thermal inertia of the corresponding location.

Figure 9 presents the THEMIS thermal inertia map for the FRT3E12 region, overlain by two HIRISE images (PSP_002176_2025_RED and ESP_026992_2025_RED). In this image, blue corresponds to low thermal inertia (more fine grained dust) and red corresponds high thermal inertia (typically more bedrock). The large red arrow (also shown in Figure 8) which indicates an area where dunes can be seen to have formed, and this area displays correspondingly low thermal inertia, and hence low grain size, as expected for aeolian material as compared to the surrounding bedrock. Figure 8a shows a CRISM spectrum (in red) taken from the area marked by the red arrow. This spectrum shows that this low thermal inertia region is covered by long wavelength-shifted olivine material. The white color at this location in Figure 8c shows this is some of the most long wavelength-shifted olivine material in the image.

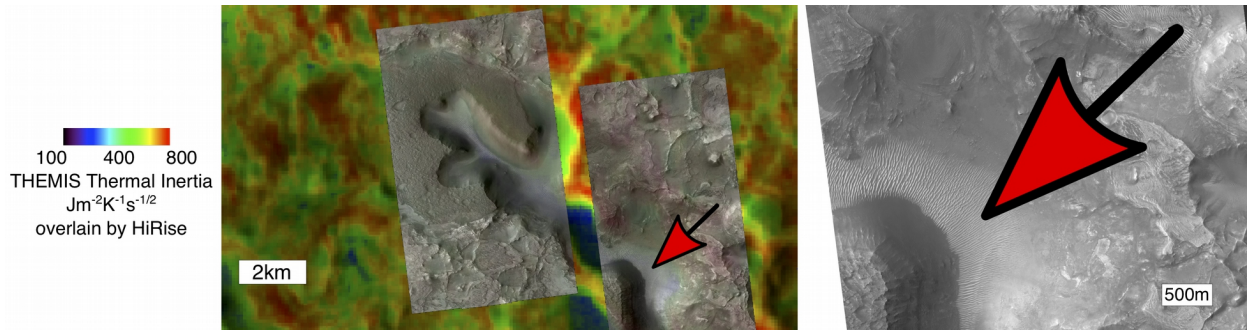

**Figure 9.** HIRISE images PSP_002176_2025_RED and ESP_026992_2025_RED overlain on THEMIS Thermal Inertia maps in the same region as CRISM image FRT3E12. Blue represents low thermal inertia, fine grain material, red represents bedrock, large grain size material. The red arrows indicate low thermal inertia but long wavelength shifted olivine material. The rightmost panel inset shows the surface is covered with linear dunes.

# 4 Discussion

## *4.1 Potential Effects on olivine 1 µm band position*

The main findings of this study rest upon apparent shifts of the central position and asymmetry of the 1 µm band complex due to olivine. This band has been documented to apparently shift to low or high wavelengths for a variety of physical scenarios (Mustard et al., 2005; Poulet et al., 2007), some of which are relevant to this study of Martian olivine composition. The olivine 1 µm band complex does not exhibit a shift in band centroid as temperature decreases to 80K (Singer and Roush, 1985), therefore we do not consider the effect of temperature changes in this study. We first examine the effect of mixing with pyroxene, then dust, then carbonates, then consider the effect of changes in grain size of the olivine sample.

### *4.1.1 Effects of intimate mixing with pyroxene*

As with any remote sensing study, we must consider the effect of contaminants and how they may affect our band fitting procedures. In order to determine the potential effects of mixing on centroid and asymmetry ranges of the 1 µm band, we applied the Asymmetric Gaussian fitting algorithm described earlier to a range of spectra that include progressively larger amounts of pyroxene mixed with olivine.

In Figure 10 we show plots of two studies of mixing of olivine and pyroxene that we obtained from RELAB. These studies were carried out by Corrigan et al. (2007) (Figure 10a) and Freeman et al. (2010) (Figure 10b). Figure 10a shows that the presence of pyroxene has the effect of moving the 1 µm band to the left (shorter wavelengths). Corrigan et al. used the San Carlos olivine which is ~Fo84.

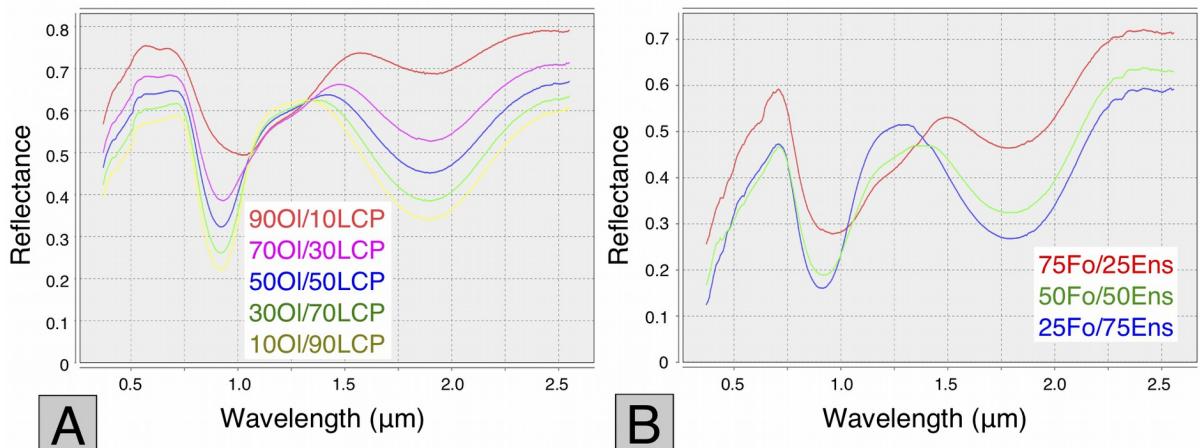

**Figure 10.** Effects on 1 µm band position of intimate mixing of olivine with pyroxene, showing a shift to shorter wavelengths with greater amounts of pyroxene. **a.)** Intimate mixing of olivine with low calcium pyroxene (LCP) from Corrigan et al. (2007) **b.)** Intimate mixing of Forsterite olivine and Enstatite pyroxene, data from Freeman et al. (2010).

In order to determine the magnitude of this effect and how it would affect our study, we first fitted the 1 μm band of the red spectrum in Figure 10a, which has the smallest amount of pyroxene (90/10). Our fit for the 90/10 mixture revealed the centroid of the 1 μm band was at 1.025 μm. In addition, the asymmetry of the 1 μm band for the 90/10 spectrum was found to be 1.255. We then carried out a fit of the blue spectrum, which is 90% pyroxene and 10% olivine. The fit for the 10/90 mixture gave a centroid of 0.926, which results in a maximal potential shift of 99.5nm for pyroxene mixing in the 90 to 10% range available with this dataset. The asymmetry for the 10/90 mixture was 1.329, for a difference of 0.074.

### *4.1.2 Effects of linear mixing with dust*

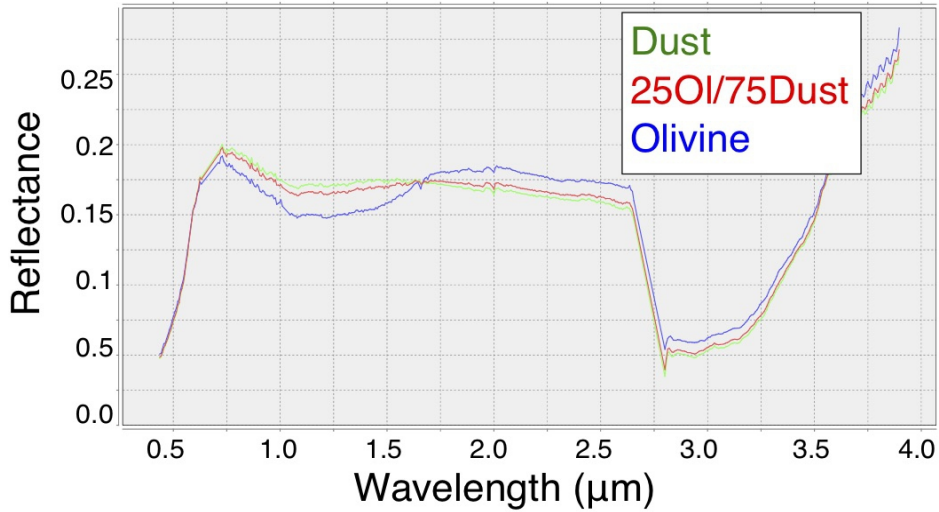

**Figure 11.** Spectral effects of mixing an example CRISM olivine spectrum with a "dusty" spectrum from the same scene in a 25% olivine/75% dust linear mixture. Locations of the spectra are shown in Figure 7.

In any CRISM pixel, it might be expected that dust or other surface component might mix with olivine elsewhere in the pixel. If this is done in a spatially separated or "checkerboard" style then the mixing can be considered to be a linear mixture of the dust spectrum with olivine. In order to assess the effect of this physical situation on the olivine 1 μm band, we have used a CRISM spectrum from FRT97E2 (see Figure 7 for continuum removed spectrum and location) with long-wavelength shifted olivine, and then mixed this in 25%, 50% and 75% amounts with a spectrally bland dust-like spectrum from elsewhere in the scene, at varying percentage amounts. The original spectrum and 75% dust mixing results are shown in Figure 11.

We then carried out an Asymmetric Gaussian fit to assess how much the mixing would shift the 1 μm band centroid. We found that the centroid of the original long wavelength olivine pixel was 1.24 μm and the centroid of the dusty spectrum is 1.09. The mixture of 75% dust gave a centroid of 1.13 μm, which is a difference of 0.11 microns.

These results demonstrate that the effect of mixing with a "dust" spectrum in the scene is to decrease the wavelength of the centroid. This strength of this effect is roughly the same as mixing with pyroxene. As for mixing with pyroxene, the effect is to move the olivine 1 μm band to shorter wavelengths. These results are summarized in Figure 15.

*4.1.3 Effects of mixing with carbonate*

We now investigate how the presence of carbonates mixed with olivine might affect the position of the 1 μm band. In order to further assess the effect of non linear (intimate) mixing with carbonate we obtained spectra from another study published in Bishop et al. (2013) and made available on RELAB. Figure 12 shows the effects of mixing of forsterite with magnesite ($MgCO_3$). As can be seen by comparing Figure 10 with Figure 12, the effect on the 1 μm band of adding magnesite to forsterite is not as profound as when pyroxene is added.

We used the Asymmetric Gaussian routine to fit the 1 μm band centroid to the extremal Fo/Mag mixtures and found the difference in centroid positions was 10.2 nm (between the Fo90/Mag10 and Fo10/Mag90 spectra). The asymmetry difference between these two extremal mixtures was 0.108. This indicates the effect of the magnesite mixing has about 10% of the effect of pyroxene mixing shifting the centroid of the 1 μm band. This carbonate mixing effect, while far weaker than the effect pyroxene mixing, is still present and moves the 1 μm band in the same direction as pyroxene (to shorter wavelengths).

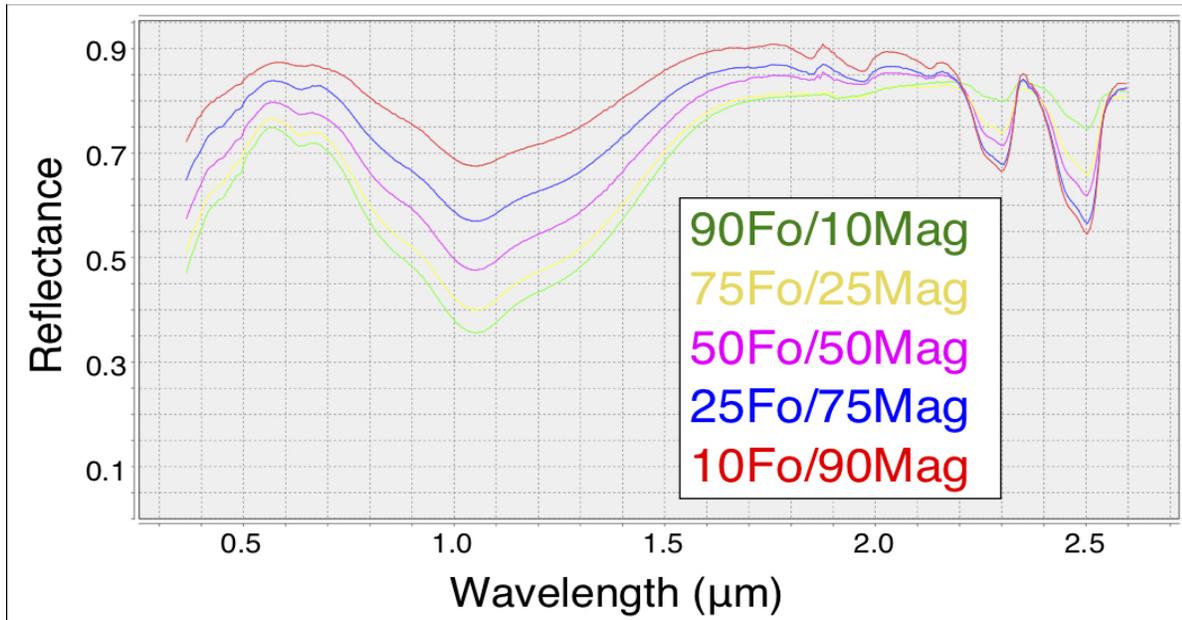

**Figure 12.** Effects of mixing olivine with Mg-carbonate, showing the shift of the 1 μm band to shorter wavelengths with greater amounts of carbonate, data are from Bishop et al. (2013).

### 4.1.4 Effects of grain size

#### 4.1.4.1 First Grain size study

In their seminal study of the olivine 1 μm band complex, King and Ridley (1987) reported centroid shifts of around 7nm (0.007 μm) for Green Sand Beach (GSB: Fo89) olivine measured with 150-250 micron grain sizes compared to <60 micron size fractions. We used the Asymmetric Gaussian model over the same sample spectra, and for the finest grain size fraction (< 60 microns), we obtained a centroid of 1.097. For the largest grain size fraction (150-200 microns) we obtained a centroid of 1.148, a difference of 50.9 nm, or +/- 25nm.

In the same manner, we calculated an error for the asymmetry parameter based on the spread of results for the four different Green Sand Beach (GSB) olivine grain sizes. The difference in asymmetry between the largest grain size and the smallest is 1.558-1.203=0.355, or +/- 0.1775. Full results are presented in the Supplementary Material section.

#### 4.1.4.2 Second Grain size study

In order to further determine the sensitivity of the olivine 1 μm band fitting to grain size effects, we used another dataset obtained from the RELAB database sourced from a previous study (Mustard and Pieters, 1989) that obtained spectra of a Fo92 olivine sample for a range of grain sizes. The sample numbers are BE-JFM-080 to 087.

Figure 13 shows the continuum removed version of the spectra of 8 different grain sizes of olivine from 25-45 to 250-500 microns (the smallest three grain sizes are almost identical). This range of grain sizes is twice as large as the King and Ridley Green Sand Beach grain size variations. We fit each of these spectra in an identical fashion with asymmetric band modeling in order to determine the variation in asymmetry and centroid. We found the centroid of the 1 μm band was 1.1063 for the 25-45 micron grain size and this changed to 1.1521 for the 250-500 micron grain size, a difference of 0.045. For the asymmetry, the 25-45 micron sample was 1.289 and the 250-500 micron asymmetry was 1.738. This means that the expected maximal error created by substituting a 25-45 micron sample for a 250-500 micron sample would be 0.045 microns, 45 nm in centroid, about the same as the error for the King and Ridley datasets in our first grain size study.

To calculate the asymmetry error, a maximal error of 0.449 is obtained between the largest and smallest datasets. This is relatively close to the error of 0.355 calculated using the King and Ridley dataset. Since the potential error estimates were relatively close, which indicates that grain size effects from these two studies are relatively congruent. However, both these datasets are limited to grain sizes less than 500 microns.

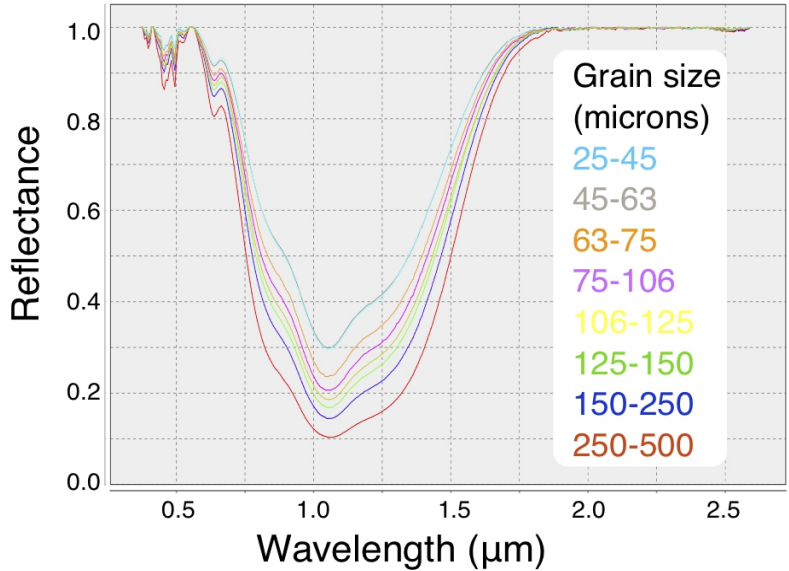

**Figure 13.** Continuum removed spectra showing the effect of grain size on the 1 μm band for the second grain size study.

*4.1.4.3 Synthetic Model Grain size study*

The King and Ridley KI olivine dataset we carried out testing with is composed of fine grained material (nearly all <160 microns). As we have previously discussed, the grain size of material on Mars is not well constrained and likely quite variable over the area of the olivine-carbonate lithology, therefore, in order to explore the full range of spectral effects of variable grain size, we have chosen to synthetically model the effects of variable grain size on the apparent band position, over the range of 70 microns to 1mm, in order to cover a wider range of grain sizes that are available in the King and Ridley dataset.

Shkuratov et al. (1999) proposed a technique for studying the effect of grain size on the 1 μm band shape of olivine using their radiative transfer approximation. This involves using their inversion equation to estimate the imaginary index of refraction and then using this index to estimate the reflectance for differing grain sizes (see in particular their Figure 11).

In this study, we have used the same approach to estimate the imaginary index of the range of olivine spectra and then fitted the resultant reflectance spectra to estimate the variations in asymmetry and centroid for a range of grain sizes and for the varying Fo# in our olivine laboratory spectra.

Figure 14 plots the relationship between the asymmetry and the 1 μm centroid position for the olivine 1 μm band, for different Fo # of olivine spectra from King and Ridley (1987). The colored lines correspond to three different average grain sizes: 70 microns, 500 microns and 1 mm. The Fo# points are labeled on the red line only, and they are transferrable to the blue and green lines on the marked points. This shows the effect of variable grain size and composition on the olivine 1μm band. We have also plotted the results of our Asymmetric Gaussian fit for CRISM FRT 3E12 (red shading) and HRL40FF (blue outline) from Figure 8 and 6 respectively.

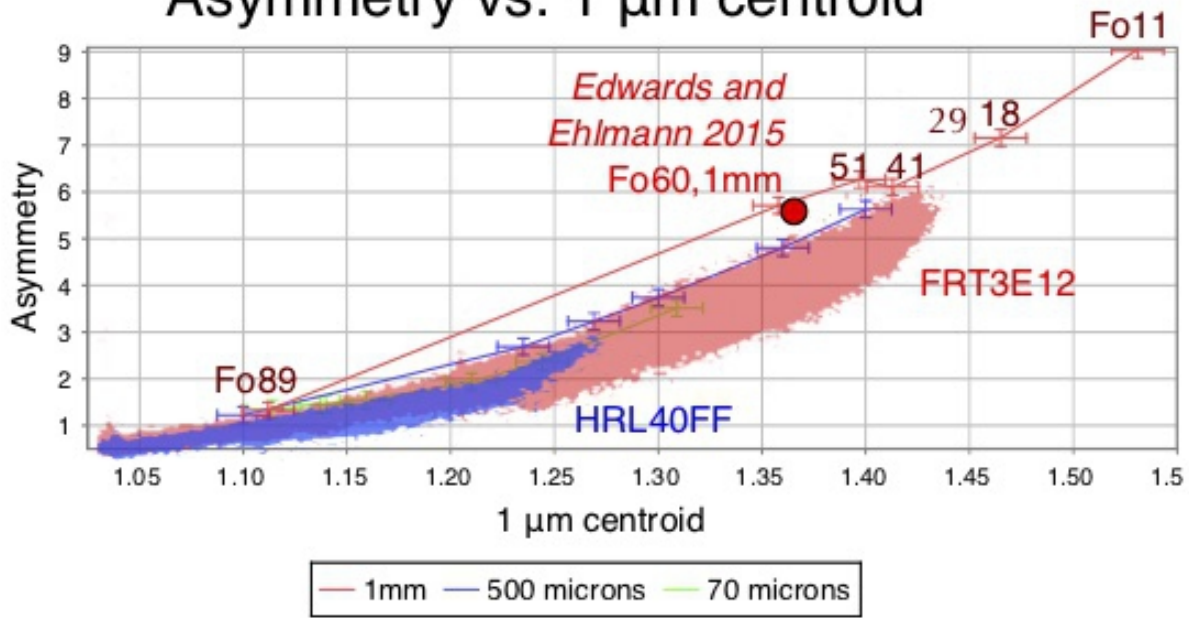

**Figure 14.** Fit of centroid versus asymmetry and olivine composition from Asymmetric Gaussian fits of the olivine 1 μm band assuming a grain size of 70, 500 and 1000 microns. Red shaded region corresponds to points in FRT3E12, blue to HRL40FF.

Known inherent errors in the Asymmetric Gaussian method in Figure 14 (as discussed in the previous section) are approximately +/- 25nm, and are primarily due to uncertainties introduced by the reliance on experimental data that introduces spread in the process.

**Bounds on composition and grain size.** Figure 14 gives us a way to estimate the bounds of the composition and grain size for the two CRISM images (these two are typical examples, see the Supplementary Material for more plots of this type for other images in the watershed). All of the bounds discussed below relate to the spectrally dominant olivine, and the caveats of Section 4.1 should be taken into account.

We have plotted the position on the plot of the recent study of Edwards and Ehlmann (2015) which carried out a Hapke fit to a CRISM spectrum in Nili Fossae (FRTC968), north of our study area, but still in the same olivine-carbonate lithological unit. They reported an olivine of 1mm grain size and Fo60 was required to fit the olivine 1 μm band adequately. Noting that their Hapke method is likely to differ from our method in a somewhat controlled sense, we plot their point result in Figure 14, and note two ways this plot allows us to constrain composition and grain size of the olivine:

**1.) Composition bounds.** The Edwards and Ehlmann spot measurement is reasonable for the location from which it was taken (it corresponds to an upper part of the red shading range of FRT3E12), however is not indicative of the full range of olivine composition for the olivine carbonate lithology. This can be seen in Figure 14 because points from FRT3E12 (red shaded in Figure 14) have a considerable tail that extends to the right of the Fo60 estimate. This indicates that although the Fo60 estimate might be a reasonable fit to the spectrum they chose (their Figure

2b), this same spectrum would could also be fitted with a smaller grain size olivine with lower Fo#. The closest point on the blue line beneath the red circle corresponds to another reasonable fit using olivine of Fo18 with 500 micron grain size, for example. This is why single point assessments are not prudent in this situation and instead an effort to place bounds is advisable.

The points plotting in the long wavelength tail in FRT3E12 corresponds to a centroid of ~1.43 μm, which intersects the red 1 mm grain size line just to the right of Fo41. Therefore, assuming a grain size of 1 mm allows us to place a lower bound of Fo40 on the composition of the olivine in FRT3E12, where it is best exposed. It should be emphasized that this is an apparent lower band, and the Fo# could be lower if some dust is present. The Fo# could also be lower if the grain size is smaller than 1mm, as discussed above and as seen from the plot.

It is worth noting that Figure 14 does not give us a way to bound the upper Fo#, because as discussed above, the effects of mixing with dust, pyroxene, carbonate will all independently shift the band position to the shorter wavelengths. This is also why we concentrate on the most extreme long wavelength shifted olivine spectra, because their shift to the long wavelengths can only be accomplished through grain size and composition.

**2.) Grain size bounds.** Figure 14 also gives us a way to place an upper and lower limit on the size of the spectroscopically dominant olivine grains in FRT3E12. We place the maximum and minimum values in two different ways.

First, the maximal value can be established by noting that in Figure 14, the red line showing the 1mm grain size family is always higher than the top values of the points for FRT3E12 (the red shaded area). FRT3E12 is the most extreme endmember image that we have analyzed to date. This strongly indicates that 1 mm grain sizes are the maximum required to explain our observations.

Second, the minimal value can be established by looking at the combined range of asymmetry and centroid values for FRT3E12. The blue line corresponding to the 500 micron grain size calculations almost runs the entire length of the observations of 3E12. This means that a lower limit of 500 microns is required to fit the range of asymmetry/centroid observations. Lower grain sizes (e.g. the green 70 micron curve) cannot explain our observations in FRT3E12. Any average olivine grain size between 500 microns and 1 mm can explain our range of observations. The lower limit can be considered stronger because lower grain sizes, mixing with dust or other components cannot explain this asymmetry/centroid range of 3E12. This does not exclude lower grain sizes in other images. Higher grain sizes might be possible if there is significant dust cover in 3E12, however we have chosen this image for its maximal range which also suggests low amounts of dust on the outcrop in this image. An overlapping image with more dust (FRTB438) is consistent with this conclusion.

Incidentally, this grain size bounds brings us close to the grain size range used by Koeppen and Hamilton (2008) in the thermal infrared study (710-1000 microns).

It should be noted that Figure 14 shows the relationship of the Asymmetric Gaussian parameter set to olivine composition is not linear. Using this scheme, it is easier to determine the composition of Fe olivine rather than Mg olivine. Therefore, the determinations of more Fe-rich compositions can be expected to be more accurate.

*4.2 Summary of potential effects on olivine 1 µm band*

Figure 15 presents a summary of the findings of our analysis of physical effects on the olivine 1 µm band. Only composition (not shown) and larger grain sizes shift the centroid to the right. We note that while other mineral phases (including pyroxene and dust as we have shown here) are no doubt present in every CRISM olivine bearing scene, when mixed with olivine, they all tend to shift the olivine band to shorter wavelengths, making olivine spectra appear more Mg-rich.

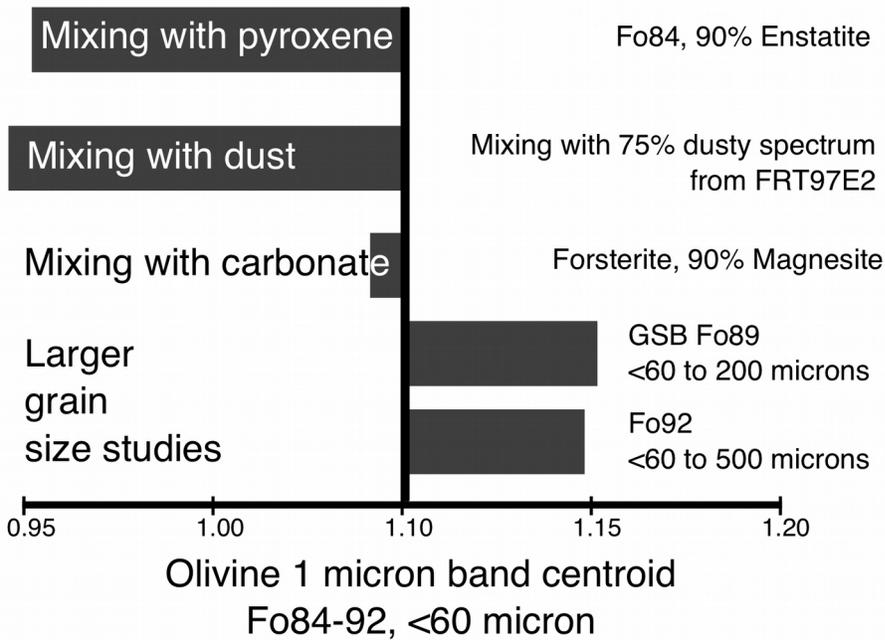

**Figure 15**. Summary of the effects of mixing and grain size on the 1 µm absorption band

*4.3 Implications for grain size effects – the thermal inertia problem*

Laboratory studies have demonstrated that larger grain sizes of olivine will shift the 1 µm band to longer wavelengths, as shown here in Figure 14 and discussed by previous workers (King and Ridley, 1987; Mustard et al., 2005; Poulet et al., 2007; Buz and Ehlmann, 2014). In attempting to constrain how this process manifests itself vis-à-vis the CRISM dataset, we have looked for evidence of large grain sizes in THEMIS thermal emission datasets, which are at just slightly poorer spatial resolution than CRISM (Fergason et al., 2006). The CRISM and THEMIS derived thermal inertia maps we have discussed herein show that there is no consistent correlation between fine grained materials and short 1 µm band centroid positions (see Figure 9). We call this the "Thermal Inertia Problem".

### 4.3.1 Summary of thermal infrared correlations with olivine 1 μm band

We firstly summarize our key observations regarding thermal inertia and the olivine 1 μm band, and then discuss two possible contributing factors that could be further explored in future studies, particularly work in situ on the olivine-carbonate lithology.

We summarize the key observations of this correlation thus: –

**1.) Dune material.** In Figure 9, at the point indicated by a red arrow, a region covered by dunes visible at HiRISE resolution displays a low thermal inertia (and interpreted low effective grain size). This dune unit also shows a strong long wavelength shift in the 1 μm band. The unit appears to consist of aeolian sand dunes created from local materials, and although the grain size has likely decreased somewhat relative to surrounding bedrock due to physical abrasion, the 1 μm band centroid remains the same, or has even slightly increased, depending on the source, which is challenging to constrain. Even if the source is not determined, the fact that fine grained dune material here corresponds to long wavelength shifted olivine cannot be explained as a grain size effect, but rather is more simply explained as a compositional effect.

**2.) Rock competence.** Geomorphologically, in many scenes, including Figure 7 and 8, the olivine-carbonate unit appears to be less competent and more prone to breakdown than surrounding geologic units, and appears to flow down slope and cover other material below it. This suggests that it is generally more friable, and easy to erode than the unit below it (the olivine-phyllosilicate bearing basement unit in Figure 8).

### 4.3.2 Potential Models for lack of correlation

**1.) Mineralogical sorting erosive aeolian regime.** It is not clear from our observations, however we consider it likely given the aeolian dunes in Figure 9 are in fact sourced from surrounding bedrock (Rogers et al., 2018). The olivine-carbonate unit does not appear to shed boulders, and appears less resistant to aeolian erosion than surrounding lithologies (such as the olivine-phyllosilicate basement unit or the mafic capping unit of the Syrtis Major lavas) (Rogers et al., 2018). When exposed, it is prone to break down, and it is easily turned into dune materials, these materials were termed "olivine bearing dunes" by Edwards and Ehlmann (2015). An important observation is that the olivine bearing dunes show no evidence for carbonate spectral signatures as compared to the surrounding carbonate and olivine bearing bedrock. We suggest that the current aeolian erosive regime is therefore wearing away more friable carbonate material, perhaps to create fine grained dust (e.g. (Bandfield et al., 2003)), and leaving intact large grains of more resistive olivine grains in a physical weathering mineralogical sorting process.

This process is consistent with proposed mineralogical fractionation of sands in Valles Marineris (Chojnacki et al., 2013) and may be a signature of density sorting, given that olivine is denser than carbonates and will be enriched in lag deposits (Fedo et al., 2015). The process is likely to be dependent on local aeolian forcings and grain cementation, and investigating whether this process is dominant in any particular aeolian bedform is well-suited to in situ investigation with a rover.

**2.) Thermal infrared skin depth.** Another contributing factor to the observed separation between large grain olivine features and thermal infrared measurements is that the THEMIS infrared wavelengths and CRISM visible-near infrared wavelengths have different penetration depths within a rock, which makes them sensitive to different mineralogical compositions. Given the fine grained nature of dust on Mars that THEMIS is sensitive to, it is possible that the variations of the grain size of the olivine is not dominating the thermal inertia measurements (i.e. changes in temperature). This could be enhanced if the spectroscopically dominant olivine is in fact a volumetrically small component of the rock under study. In the aeolian weathering scenario we have outlined above, if the majority of the dust is sourced from the non-olivine grains, this will build up over time as a fine grained lag and control the temperature of the outcrop, and the olivine grains will thereby not participate in the temperature regime of the outcrop.

## 5. Implications and formation models

Thus far in this study, we have shown a correlation between olivine 1 μm bands shifted to long wavelengths and carbonates, and we have analysed the possible physical effects that might shift the 1 μm band. We have demonstrated that the saturation of the olivine 1 μm band occurs in some regions, and there is variability of this saturation throughout the olivine-carbonate lithology. In our three study areas (Figure 6, 7 and 8) we have demonstrated that carbonate is not associated with the most saturated olivine, or the least saturated, but is instead associated with intermediate saturations (those characterised by a centroid of 1.2-1.3 μm, Figure 6b, 7b, 8b).

We have shown that the most saturated olivine signatures do not correlate with carbonates (Figure 8b), *potentially* indicating these locations are the least reacted olivines in the lithology and an extensive carbonation event has not affected them.

We now highlight three key science questions that have arisen from the results of this study of the olivine-carbonate lithology.

### *5.1. What controlled the composition of olivine in the olivine-carbonate lithology?*

The first question posed by our findings is perhaps the most obvious: what controls the olivine composition of the olivine-carbonate lithology? If the origin of the observed olivine-carbonate lithology is volcanic (with some later in-situ aeolian modification by density sorting (Fedo et al., 2015)), the original variation in Mg-Fe geochemistry of Nili Fossae olivines could have been: 1.) controlled by ponding and fractionation in a magma chamber isolated from the mantle (Filiberto and Dasgupta, 2011), and/or 2.) reflect cooling and less vigorously mixed martian mantle (Filiberto and Dasgupta, 2015). If a volcanic origin scenario associated with relatively cool volcanic flow is indeed true, then the association of carbonates with ~Fo40 olivine again begs the key question:- how was a relatively cooler lava package able to drive serpentinization reactions to induce carbonate formation? Potential answers to this question may begin with a carbonate rich mantle source, which was more common the early Noachian (Grott et al., 2011), as proposed for the Wishstone class of basalts at Gusev crater (Usui et al., 2008). The relatively low temperature magmas at Nili Fossae may be driven by local crustal thinning due to the Isidis impact (Kiefer, 2005). Obviously, in situ analyses of these units will shed light on these hypotheses.

Olivine dominated basalts from Gusev crater have been analyzed using the MER Rover Spirit's instrumentation, and found to have Fo40-60 composition (McSween et al., 2006). It is therefore important to better understand the composition of the olivine in the olivine-carbonate unit, which is likely the best exposed oldest olivine accessible to surface exploration. The co-evolution of the Martian crust and mantle has been constrained by numerical cooling models of the interior assuming a stagnant lid planet, and the variation in LCP to HCP composition of pyroxene has been found to match the observations at the Noachian/Heseperian boundary (Baratoux et al., 2013). Olivine composition was not modelled in that study.

*In situ* analysis of the Jezero olivine-carbonate lithologies to ground truth the maps created here and discover whether there are further geochemical or mineralogical differences between the lithologies is likely to have implications for the formation of the entire mafic surface of Mars, as suggested by the global studies of Poulet et al. (2007), Ody et al. (2013) and Clenet et al. (2013). These global studies found (Type 2) Fe-olivine in a restricted region at Nili Fossae, and thus far, the best global manifestations of carbonates, as we have shown here for the first time, these unusual units are not just in the same region, but in the same lithological unit (at the CRISM scale).

### *5.2. Why is the olivine-carbonate lithology only partially carbonatized?*

The second question arising from our study is: why is the olivine-carbonate lithology only partially carbonatized? We will split our discussion into two types of carbonization mechanisms associated with olivine emplacement: 1) contemporaneous emplacement, and 2) a distinct carbonation event following emplacement.

#### *5.2.1 Contemporaneous carbonation event*

If the origin of the olivine-carbonate lithology is largely volcanic, it is conceivable that volcanic fluids bearing $CO_2$ from the mantle (Grott et al., 2011) were erupted with the olivine and were subsequently serpentinized into carbonate rich inclusions within the erupted lava. This process may also have happened as part of an ash fall deposit, as has been documented in smaller terrestrial systems (Willcox et al., 2015). The source of the fluids would have been connate or juvenile water sourced from a $CO_2$ rich region of the mantle. The heat of the lava would have driven hydrothermal serpentinization reactions, (typically <400ºC) and would have cooled relatively quickly in order to preserve the original volcanic or pyroclastic olivine (Klein and Garrido, 2011; Klein and McCollom, 2013).

Typical terrestrial studies of ultramafic serpentinisation examine the $Mg-SiO_2-H_2O$ sequence (Moody, 1976), however some serpentinisation of Fe-bearing olivine has been recently investigated (Klein et al., 2013). Klein et al. modeled the reactions of a complete range of Forsterite-Fayalite olivine at temperatures <400C in the presence of water to form serpentine +/- brucite +/- talc +/- magnetite, and found that fayalite is more productive than forsterite under the modeled conditions. They did not simultaneously model the production of carbonate for the forsterite-fayalite sequence.

In these scenarios, the partial carbonation of the olivine lithology might be explained by the restriction of $CO_2$ rich fluids to variable porosity or fractures in the rock that restricted fluid pathways, resulting in an uneven distribution of carbonate throughout the unit. The fact that olivine is still present in these rocks is evidence that of the relatively feeble nature of the hydrothermal system that drove the serpentinization. In more persistent and pervasive terrestrial serpenitinization systems, particularly in submarine environments, it is common for the olivine to be completely replaced.

*5.2.2 Post-emplacement Carbonation event*

Scenarios of secondary carbonate emplacement following deposition of the olivine rich layers are also possible. If the atmosphere was thicker when the olivine layers were deposited, it is possible that the carbonate may have been emplaced under a greenhouse atmosphere (Pollack et al., 1987), and the olivine-carbonate lithology may thus be a record of an early Martian carbon cycle. In this scenario, one might consider the variable carbonation problematic, unless it is due to variable exposure of the olivine-carbonate lithology, which has not yet been uncovered.

Investigating the potential of this scenario, Edwards and Ehlmann (2015) studied the process of low temperature carbonation of olivine in the subsurface, using the model of van Berk and Fu (2011). They calculated that the amount of olivine-carbonate currently exposed at Nili Fossae is insufficient to generate a greenhouse atmosphere sufficient to warm the surface above freezing.

It is also possible that the carbonation was emplaced during the period when a lake was present at Jezero crater (Horgan and Anderson, 2018). The presence of similar olivine-carbonate lithologies in the Jezero crater (Figure 6) and watershed (Figure 7) is problematic for this scenario, although if it can be shown that the crater rim carbonates have a separate origin to the olivine-carbonate lithology at Jezero, this scenario would be more likely.

*5.3. What is the relationship between carbonates and Mg-phyllosilicate?*

The third question our study raises is: what is the relationship of carbonates to associated Mg-phyllosilicate signatures and the hydration event(s) that emplaced them? Were the hydration and carbonation events contemporaneous? This study has not concentrated on origins of the phyllosilicates or their compositions, even though they are ubiquitous and variable in the olivine-carbonate lithology. *In situ* investigation of the mineralogy associated with the carbonate layer, including the identification of specific Mg-phyllosilicates such as smectite (Bishop et al., 2008), chlorite (Viviano et al., 2013), saponite (Ehlmann et al., 2009), serpentine (Ehlmann et al., 2009; Brown et al., 2010; Amador et al., 2018) and/or talc (Brown et al., 2010; Viviano et al., 2013), will help reveal the alteration conditions and temperature and pressure conditions that accompanied the hydration event, and may determine whether it was associated with a serpentinization, sedimentary or leaching process. The formation conditions are crucial because the presence of talc-carbonate resulting from the carbonation of serpentine has been examined in Earth analogs in terrestrial greenstone belts such as the Pilbara in Western Australia (Brown et al., 2005, 2006; Brown, 2006), where talc-bearing komatiite cumulate units of the Dresser Formation overlie the siliceous, stromatolite-bearing Strelley Pool Chert unit (Van Kranendonk et al., 2008). An in situ investigation of the Mg-phyllosilicate mineralogy and the nature of the hydration event is therefore a critical task in understanding the astrobiological potential of the carbonate and phyllosilicate deposits at Nili Fossae.

*5.4 Proposed Olivine-Carbonate lithology Formation Scenarios*

Table 2 presents a summary of potential formation scenarios we are presently aware of that may have played a role in the emplacement and/or alteration of the olivine-carbonate lithology in the Jezero crater region. Each of the formation and alteration scenarios in Table 2 present a plausible formation mechanism for the emplacement of carbonate at Nili Fossae, and potentially the accompanying olivine and associated Mg-phyllosilicates. We have made a qualitative assessment of the ability of each scenario to address the findings of this study, based on current descriptions in the literature. At this time we do not have the required data to conclude which scenario is correct, and/or whether multiple scenarios were at play. However, in the last column of the table, we have provided testable hypotheses and/or observables for each scenario that could be used to evaluate each mechanism with future *in situ* exploration.

| Scenario | 1. Olivine Emplacement | 2. Partial carbonitization | 3. Associated Mg-phyllo | In situ testable hypothesis or observable |
|---|---|---|---|---|
| Volcanic succession on early Martian crust (Hamilton and Christensen, 2005; Tornabene et al., 2008), by dyke driven volcanism (Bramble et al., 2017) with accompanying deuteric alteration; serpentinization reactions driven by heat of volcanic emplacement (Brown et al., 2010; Viviano et al., 2013) | √√√ | √√√ | √√√ | Lava flow units in stratigraphic section |
| Impact-driven hydrothermal activity by a melt sheet (Hoefen et al., 2003; Mustard et al., 2007, 2009) or by pyroclastic ash fall (Kremer et al., 2018); serpentinization reactions driven by hydrothermal activity from heat of impact (Osinski et al., 2013) | √√ | √√√ | √√ | Superposed impact melt sheet or pyroclastic ash unit in stratigraphic section |
| Subsurface alteration under thicker $CO_2$ atmosphere; serpentinization reactions driven by diagenesis and upper crustal hydrothermal processes (van Berk and Fu, 2011; Edwards and Ehlmann, 2015) | | √√√ | √√√ | Indicators of pedogenic alteration in carbonate and layering as predicted in (van Berk and Fu, 2011). |
| Hydrothermal alteration in thermal springs environment (Walter and Des Marais, 1993) or alteration of volcanic tephra by ephemeral waters (Ruff et al., 2014) | | √√√ | √√ | Mineralogical and physical evidence of tephra-like deposits showing hydrothermal alteration |
| Deep Subsurface reservoir of carbonate exposed by meteor impact (Michalski and Niles, 2010; Glotch and Rogers, 2013) | | √√√ | √√ | Layering, exposure in deep crater walls or peaks |
| Cold ophiolite-hosted serpentinization, as in the terrestrial analogs in California (Campbell et al., 2002; Schulte et al., 2006) or the Oman ophiolite (Paukert et al., 2012) | | √√ | √ | Low temperature serpentinization minerals (See discussion) |

| Scenario | | | | |
|---|---|---|---|---|
| Low temperature leaching, as in terrestrial analog of Antarctic carbonate rinds (Doran et al., 1998; Salvatore et al., 2013) and Mojave desert carbonate rinds (Bishop et al., 2011) | | √√ | √ | Carbonate in surface rinds |
| Precipitation of carbonate directly into shoreline of shallow lake (Horgan and Anderson, 2018) or dry lake (Baldridge et al., 2009; Marion et al., 2009) or marine basin, includes scenarios of Noachian Martian ocean preserved at Nili Fossae (Russell et al., 2014) | | √ | √√ | Marginal carbonates, Carbonate reefs, microbiolites, stromatolites exhalative/smoker structures if deep deposits |
| Hydrothermal formation of carbonates and clays from an olivine bearing Martian basalt under a thick $CO_2$ atmosphere (Pollack et al., 1987; Dehouck et al., 2014) or under a high pressure and temperature atmosphere (Cannon et al., 2017) | √√√ | √√ | √√√ | High temperature or pressure mineral phases (see Discussion) |
| Deposition of olivine/carbonate sediment in a large aeolian dune field, such as the lower unit of the Burns Formation (Grotzinger et al., 2005) | √√ | √ | √ | Aeolian sedimentary features in the carbonate |

**Table 2.** Possible Formation scenarios, science questions arising from this study and testable predictions for Jezero olivine-carbonate lithology. No ticks indicates the scenario cannot address this question, one tick indicates the scenario might account for this question, two ticks indicates it partially deals with the question, and three ticks means it specifically addresses the question.

# 6 Conclusions

This study has provided four new insights into the Nili Fossae olivine-carbonate lithology.

1.) We have established that within the olivine-carbonate lithology, there is band saturation of the olivine 1 μm band complex. Variability of this saturation effect has enabled us to place limits on the grain size of olivine controlling the band shape. We have shown using an Asymmetric Gaussian fit in three key regions of Nili Fossae that the centroid and asymmetry that the presence of at least 500 micron grain size olivine is required, and at most 1 mm grains can be accomodated. Additionally, if the 1mm grain size is assumed, we have shown evidence that these can be fitted with a Fo40 olivine and that some observations are so saturated they cannot be fit by 1mm grain size, Fo60 olivine.

2.) We have established that the portion of the Nili Fossae olivine characterized in the VNIR as Fe-olivine (or large grain size material) by Mustard et al. (2005), Poulet et al. (2007), mapped as Type II olivine by Ody et al. (2013) and mapped by Clenet et al. (2013), is in fact spatially associated with the carbonates discovered at Nili Fossae by Ehlmann et al. (2008b). We have established that this association is only with intermediate saturation olivine units, and not with maximally or minimally saturated olivines.

3.) We have studied thermal inertia maps from THEMIS to better characterize the olivine-carbonate lithology. We have showed that the olivine-carbonate 1 μm centroid does not display a reliable correlation with thermal inertia, and we have posited several potential reasons for this.

We have demonstrated that obscuration by dust and other components can shift the olivine band to shorter wavelengths, however it cannot shift the band to longer wavelengths or explain our observations of band saturation. We have demonstrated that the observations we have made of olivine band saturation and centroid requires a combination of large grain size and Fe-rich composition. We fully realize our demonstration and the non linear curves we have derived are dependent on a particular and relatively simple radiative transfer model, however there are good physical reasons to believe these large grain size-saturation trends will hold. For example, band saturation occurs for large grain sized water ice in the summer exposures of the north polar residual ice cap (Langevin et al., 2005; Brown et al., 2016).

This study has therefore strengthened our geological understanding of two previously heavily studied but heretofore unlinked key units of the Nili Fossae region. It has placed constraints on the grain size and composition of the olivine-carbonate lithology using the centroid position and asymmetry of the 1 μm olivine band complex. Carbonate has been shown to only occur in association with intermediate saturation olivine spectra. In addition, we have discussed several olivine-carbonate formation pathways and how this correlation between olivine and carbonate is expected to impact previously proposed formation scenarios.

**Acknowledgments, Samples, and Data**

All CRISM data used can be obtained from the Planetary Data System (PDS) Geoscience Node (http://ode.rsl.wustl.edu/mars). We would also like to thank Keith Putirka for discussions on volcanic and early Martian core conditions. We would like to thank Robin Fergason for her help interpreting the THEMIS TI dataset. We would also like to thank Scott Murchie for reviewing a draft of this paper and Frank Seelos and the CRISM team at JHUAPL for their untiring work to produce a fantastic dataset.

AJB conducted this study with the support of the NASA Astrobiology Institute, (Grant# NNX15BB01A) and CEV and AJB with coverage from NASA MDAP (Grant# NNX16AJ48G). TAG acknowledges support from the CRISM team through a subcontract from the Johns Hopkins University Applied Physics Lab.

**References**

Amador, E.S., Bandfield, J.L., Thomas, N.H., 2018. A search for minerals associated with serpentinization across Mars using CRISM spectral data. Icarus 311, 113–134. https://doi.org/10.1016/j.icarus.2018.03.021

Baldridge, A.M., Hook, S.J., Crowley, J.K., Marion, G.M., Kargel, J.S., Michalski, J.L., Thomson, B.J., de Souza Filho, C.R., Bridges, N.T., Brown, A.J., 2009. Contemporaneous deposition of phyllosilicates and sulfates: Using Australian acidic saline lake deposits to describe geochemical variability on Mars. Geophys. Res. Lett. 36.

Bandfield, J.L., Glotch, T.D., Christensen, P.R., 2003. Spectroscopic identification of carbonate minerals in the martian dust. Science 301, 1084–1087.

Baratoux, D., Toplis, M.J., Monnereau, M., Sautter, V., 2013. The petrological expression of early Mars volcanism. J. Geophys. Res. Planets 118, 59–64.

Bishop, J.L., Lane, M.D., Dyar, M.D., Brown, A.J., 2008. Reflectance and emission spectroscopy of four groups of phyllosilicates: smectites, kaolinite-serpentines, chlorites and micas. Clay Miner. 43, 35–54.

Bishop, J.L., Perry, K.A., Darby Dyar, M., Bristow, T.F., Blake, D.F., Brown, A.J., Peel, S.E., 2013. Coordinated spectral and XRD analyses of magnesite-nontronite-forsterite mixtures and implications for carbonates on Mars. J. Geophys. Res. Planets 635–650.

Bishop, J.L., Schelble, R.T., McKay, C.P., Brown, A.J., Perry, K.A., 2011. Carbonate rocks in the Mojave Desert as an analogue for Martian carbonates. Int. J. Astrobiol. 10, 349–358. https://doi.org/10.1017/S1473550411000206
Bramble, M.S., Mustard, J.F., Salvatore, M.R., 2017. The geological history of Northeast Syrtis Major, Mars. Icarus 293, 66–93. https://doi.org/10.1016/j.icarus.2017.03.030
Brown, Adrian J., 2006. Spectral Curve Fitting for Automatic Hyperspectral Data Analysis. IEEE Trans. Geosci. Remote Sens. 44, 1601–1608. https://doi.org/10.1109/TGRS.2006.870435
Brown, A.J., 2006. Hyperspectral Mapping of Ancient Hydrothermal Systems (PhD). Macquarie University, Sydney, N.S.W.
Brown, A.J., Calvin, W.M., Becerra, P., Byrne, S., 2016. Martian north polar cap summer water cycle. Icarus 277, 401–415. https://doi.org/10.1016/j.icarus.2016.05.007
Brown, A.J., Cudahy, T.J., Walter, M.R., 2006. Hydrothermal alteration at the Panorama Formation, North Pole Dome, Pilbara Craton, Western Australia. Precambrian Res. 151, 211–223. https://doi.org/10.1016/j.precamres.2006.08.014
Brown, A.J., Hook, S.J., Baldridge, A.M., Crowley, J.K., Bridges, N.T., Thomson, B.J., Marion, G.M., de Souza Filho, C.R., Bishop, J.L., 2010. Hydrothermal formation of Clay-Carbonate alteration assemblages in the Nili Fossae region of Mars. Earth Planet. Sci. Lett. 297, 174–182. https://doi.org/10.1016/j.epsl.2010.06.018
Brown, A.J., Walter, M.R., Cudahy, T.J., 2005. Hyperspectral Imaging Spectroscopy of a Mars Analog Environment at the North Pole Dome, Pilbara Craton, Western Australia. Aust. J. Earth Sci. 52, 353–364. https://doi.org/10.1080/08120090500134530
Burns, R.G., 1970. Crystal field spectra and evidence of cation ordering in olivine minerals. Am. Mineral. 55, 1608–1632.
Burns, R.G., Huggins, F.E., 1972. Cation determination curves for Mg-Fe-Mn olivines from vibrational spectra. Am. Mineral. 57, 967–985.
Buz, J., Ehlmann, B.L., 2014. Effects of grain size on the reflectance spectroscopy of olivine in the VIS-NIR and the derivation of olivine composition using Modified Gaussian Modelling. Presented at the LPSC, LPI, Houston, TX, p. #2810.
Campbell, K.A., Farmer, J.D., Des Marais, D., 2002. Ancient hydrocarbon seeps from the Mesozoic convergent margin of California: carbonate geochemistry, fluids and palaeoenvironments. Geofluids 2, 63–94.
Cannon, K., Parman, S., Mustard, J.F., 2017. Primordial clays on Mars formed beneath a steam or supercritical atmosphere. Nature 552, 88–91. https://doi.org/doi:10.1038/nature24657
Carter, J., Poulet, F., 2012. Orbital identification of clays and carbonates in Gusev Crater. Icarus 219.
Chojnacki, M., Burr, D.M., Moersch, J.E., 2013. Valles Marineris Dune Fields As Compared With Other Martian Populations: Diversity Of Dune Compositions, Morphologies, And Thermophysical Properties. Icarus.
Christensen, P.R., Engle, E., Anwar, S., Dickenshied, S., Noss, D., Gorelick, N., Weiss-Malik, M., 2009. JMARS - A Planetary GIS. AGU Fall Meet. Abstr. 22, IN22A-06.
Christensen, P.R., Ruff, S.W., Fergason, R.L., Knudson, A.T., Anwar, S., Arvidson, R.E., Bandfield, J.L., Blaney, D.L., Budney, C., Calvin, W.M., Glotch, T.D., Golombek, M.P., Gorelick, N., Graff, T.G., Hamilton, V.E., Hayes, A., Johnson, J.R., McSween, H.Y., Jr., Mehall, G.L., Mehall, L.K., Moersch, J.E., Morris, R.V., Rogers, A.D., Smith, M.D., Squyres, S.W., Wolff, M.J., Wyatt, M.B., 2004. Initial Results from the Mini-TES Experiment in Gusev Crater from the Spirit Rover. Science 305, 837–842.
Clark, R.N., Swayze, G.A., Wise, R., Livo, E., Hoefen, T., Kokaly, R., Sutley, S.J., 2007. USGS digital spectral library splib06a: http://speclab.cr.usgs.gov/spectral.lib06., Digital Data Series 231. U.S. Geological Survey.
Clenet, H., Pinet, P., Ceuleneer, G., Daydou, Y., Heuripeau, F., Rosemberg, C., Bibring, J.-P., Bellucci, G., Altieri, F., Gondet, B., the, O.T., 2013. A systematic mapping procedure based on the Modified Gaussian Model to characterize magmatic units from olivine/pyroxenes mixtures: Application to the Syrtis Major volcanic shield on Mars. J. Geophys. Res. Planets n/a-n/a.
Clenet, H., Pinet, P., Daydou, Y., Heuripeau, F., Rosemberg, C., Baratoux, D., Chevrel, S.D., 2011. A new systematic approach using the Modified Gaussian Model: Insight for the characterization of chemical composition of olivines, pyroxenes and olivine–pyroxene mixtures. Icarus 213, 404–422.
Comer, R.P., Solomon, S.C., Head, J.W., 1985. Mars: Thickness of the lithosphere from the tectonic response to volcanic loads. Rev. Geophys. 23, 61–92. https://doi.org/10.1029/RG023i001p00061

Corrigan, C.M., McCoy, T.J., Sunshine, J.M., Bus, S.J., Gale, A., 2007. Does Spectroscopy Provide Evidence for Widespread Partial Melting of Asteroids?: I. Pyroxene Compositions. Presented at the LPSC XXXVIII, LPI, Houston, TX, p. abstract 1463.

Dehouck, E., Gaudin, A., Mangold, N., Lajaunie, L., Dauzères, A., Grauby, O., Le Menn, E., 2014. Weathering of olivine under $CO_2$ atmosphere: A martian perspective. Geochim. Cosmochim. Acta 135, 170–189. https://doi.org/10.1016/j.gca.2014.03.032

Doran, P.T., Wharton, R.A., Des Marais, D.J., McKay, C.P., 1998. Antarctic paleolake sediments and the search for extinct life on Mars. J. Geophys. Res.-Planets 103, 28481–28493.

Dreibus, G., Wänke, H., 1985. Mars, A Volatile-Rich Planet. Meteoritics 20, 367–381.

Edwards, C.S., Ehlmann, B.L., 2015. Carbon sequestration on Mars. Geology 43, G36983.1. https://doi.org/10.1130/G36983.1

Ehlmann, B.L., Mustard, J.F., Fassett, C.I., Schon, S.C., Head Iii, J.W., Des Marais, D.J., Grant, J.A., Murchie, S.L., 2008a. Clay minerals in delta deposits and organic preservation potential on Mars. Nat. Geosci. 1, 355–358. https://doi.org/10.1038/ngeo207

Ehlmann, B.L., Mustard, J.F., Murchie, S.L., Poulet, F., Bishop, J.L., Brown, A.J., Calvin, W.M., Clark, R.N., Marais, D.J.D., Milliken, R.E., Roach, L.H., Roush, T.L., Swayze, G.A., Wray, J.J., 2008b. Orbital Identification of Carbonate-Bearing Rocks on Mars. Science 322, 1828–1832. https://doi.org/10.1126/science.1164759

Ehlmann, B.L., Mustard, J.F., Swayze, G.A., Clark, R.N., Bishop, J.L., Poulet, F., Des Marais, D.J., Roach, L.H., Milliken, R.E., Wray, J.J., Barnouin-Jha, O., Murchie, S.L., 2009. Identification of hydrated silicate minerals on Mars using MRO-CRISM: Geologic context near Nili Fossae and implications for aqueous alteration. J. Geophys. Res. 114, doi://10.1029/2009JE003339.

Elkins-Tanton, L.T., Hess, P.C., Parmentier, E.M., 2005. Possible formation of ancient crust on Mars through magma ocean processes. J. Geophys. Res. Planets 110, E12S01. https://doi.org/10.1029/2005JE002480

Fassett, C.I., Head, J.W., 2008. Valley network-fed, open-basin lakes on Mars: Distribution and implications for Noachian surface and subsurface hydrology. Icarus 198, 37–56. https://doi.org/10.1016/j.icarus.2008.06.016

Fassett, C.I., Head, J.W., 2005. Fluvial sedimentary deposits on Mars: Ancient deltas in a crater lake in the Nili Fossae region. Geophys. Res. Lett. 32, L14201. https://doi.org/10.1029/2005GL023456

Fedo, C.M., McGlynn, I.O., McSween, H.Y., 2015. Grain size and hydrodynamic sorting controls on the composition of basaltic sediments: Implications for interpreting martian soils. Earth Planet. Sci. Lett. 423, 67–77. https://doi.org/10.1016/j.epsl.2015.03.052

Fergason, R.L., Christensen, P.R., Kieffer, H.H., 2006. High-resolution thermal inertia derived from the Thermal Emission Imaging System (THEMIS): Thermal model and applications. J. Geophys. Res. 111, doi:10.1029/2006JE002735.

Filiberto, J., Dasgupta, R., 2015. Constraints on the depth and thermal vigor of melting in the Martian mantle. J. Geophys. Res. Planets 2014JE004745. https://doi.org/10.1002/2014JE004745

Filiberto, J., Dasgupta, R., 2011. $Fe^{2+}$-Mg partitioning between olivine and basaltic melts: Applications to genesis of olivine-phyric shergottites and conditions of melting in the Martian interior. Earth Planet. Sci. Lett. 304, 527–537.

Freeman, W., Bishop, J., Marchis, F., Emery, J., Reiss, A.E., Hiroi, T., Navascués, D.B. y., Shaddad, M.H., Jenniskens, P., 2010. Investigation of the Origin of 2008TC3 Through Spectral Analysis of F-type Asteroids and Lab Spectra of Almahata Sitta and Mineral Mixtures. Bull. Am. Astron. Soc. 42, 13.31.

Frey, H., 2008. Ages of very large impact basins on Mars: Implications for the late heavy bombardment in the inner solar system. Geophys. Res. Lett. 35. https://doi.org/10.1029/2008GL033515

Glotch, T.D., Rogers, A.D., 2013. Evidence for magma-carbonate interaction beneath Syrtis Major, Mars. J. Geophys. Res. Planets n/a-n/a.

Goudge, T.A., Milliken, R.E., Head, J.W., Mustard, J.F., Fassett, C.I., 2017. Sedimentological evidence for a deltaic origin of the western fan deposit in Jezero crater, Mars and implications for future exploration. Earth Planet. Sci. Lett. 458, 357–365. https://doi.org/10.1016/j.epsl.2016.10.056

Goudge, T.A., Mustard, J.F., Head, J.W., Fassett, C.I., Wiseman, S.M., 2015. Assessing the Mineralogy of the Watershed and Fan Deposits of the Jezero Crater Paleolake System, Mars. J. Geophys. Res. Planets 2014JE004782. https://doi.org/10.1002/2014JE004782

Greenwood, H.J., 1967. Mineral equilibria in the system $MgO-SiO_2-H_2O-CO_2$. Res. Geochem. 2, 542–567.

Grott, M., Breuer, D., 2008. The evolution of the martian elastic lithosphere and implications for crustal and mantle rheology. Icarus 193, 503–515. https://doi.org/10.1016/j.icarus.2007.08.015
Grott, M., Morschhauser, A., Breuer, D., Hauber, E., 2011. Volcanic outgassing of CO2 and H2O on Mars. Earth Planet. Sci. Lett. 308, 391–400.
Grotzinger, J.P., Arvidson, R.E., Bell, J.F., Calvin, W., Clark, B.C., Fike, D.A., Golombek, M., Greeley, R., Haldemann, A., Herkenhoff, K.E., Jolliff, B.L., Knoll, A.H., Malin, M., McLennan, S.M., Parker, T., Soderblom, L., Sohl-Dickstein, J.N., Squyres, S.W., Tosca, N.J., Watters, W.A., 2005. Stratigraphy and sedimentology of a dry to wet eolian depositional system, Burns formation, Meridiani Planum, Mars. Earth Planet. Sci. Lett., Sedimentary Geology at Meridiani Planum, Mars 240, 11–72. https://doi.org/10.1016/j.epsl.2005.09.039
Hamilton, V.E., Christensen, P.R., 2005. Evidence for extensive, olivine-rich bedrock on Mars. Geology 33, 433–436.
Hemley, J.J., Montoya, J.W., Shaw, D.R., Luce, R.W., 1977. Mineral equilibria in the system MgO-SiO2-H2O system: II. Talc-antigorite-forsterite-enstatite stability relations and some general implications in the system. Am. J. Sci. 277, 353–383.
Hiesinger, H., Head, J.W., 2004. The Syrtis Major volcanic province, Mars: Synthesis from Mars Global Surveyor data. J. Geophys. Res. 109, 10.1029/2003JE002143.
Hoefen, T.M., Clark, R.N., Bandfield, J.L., Smith, M.D., Pearl, J.C., Christensen, P.R., 2003. Discovery of Olivine in the Nili Fossae Region of Mars. Science 302, 627–630.
Hoke, M.R.T., Hynek, B.M., 2009. Roaming zones of precipitation on ancient Mars as recorded in valley networks. J. Geophys. Res. 114.
Horgan, B.H.N., Anderson, R.B., 2018. Possible Lacustrine Carbonates in Jezero Crater, Mars - A Candidate Mars 2020 Landing Site. Presented at the LPSC XXXXIX, LPI, Houston, Tx, p. #1749.
Irwin, R.P., Howard, A.D., Craddock, R.A., Moore, J.M., 2005. An intense terminal epoch of widespread fluvial activity on early Mars: 2. Increased runoff and paleolake development. J. Geophys. Res. Planets 110, E12S15. https://doi.org/10.1029/2005JE002460
Isaacson, P.J., Klima, R.L., Sunshine, J.M., Cheek, L.C., Pieters, C.M., Hiroi, T., Dyar, M.D., Lane, M., Bishop, J., 2014. Visible to near-infrared optical properties of pure synthetic olivine across the olivine solid solution. Am. Mineral. 99, 467–478. https://doi.org/10.2138/am.2014.4580
Isaacson, P.J., Pieters, C.M., Besse, S., Clark, R.N., Head, J.W., Klima, R.L., Mustard, J.F., Petro, N.E., Staid, M.I., Sunshine, J.M., Taylor, L.A., Thaisen, K.G., Tompkins, S., 2011. Remote compositional analysis of lunar olivine-rich lithologies with Moon Mineralogy Mapper (M3) spectra. J. Geophys. Res. Planets 116, E00G11. https://doi.org/10.1029/2010JE003731
Kiefer, W.S., 2005. Buried mass anomalies along the hemispheric dichotomy in eastern Mars: Implications for the origin and evolution of the dichotomy. Geophys. Res. Lett. 32. https://doi.org/10.1029/2005GL024260
King, T.V.V., Ridley, W.I., 1987. Relation of the spectroscopic reflectance of olivine to mineral chemistry and some remote sensing implications. J. Geophys. Res. Solid Earth 92, 11457–11469. https://doi.org/10.1029/JB092iB11p11457
Klein, F., Bach, W., McCollom, T.M., 2013. Compositional controls on hydrogen generation during serpentinization of ultramafic rocks. Lithos 178, 55–69.
Klein, F., Garrido, C.J., 2011. Thermodynamic constraints on mineral carbonation of serpentinized peridotite. Lithos 126, 147–160. https://doi.org/10.1016/j.lithos.2011.07.020
Klein, F., McCollom, T.M., 2013. From serpentinization to carbonation: New insights from a CO2 injection experiment. Earth Planet. Sci. Lett. 379, 137–145. https://doi.org/10.1016/j.epsl.2013.08.017
Koeppen, W.C., Hamilton, V.E., 2008. Global distribution, composition, and abundance of olivine on the surface of Mars from thermal infrared data. J. Geophys. Res. Planets 113, https://doi.org/10.1029/2007JE002984.
Kremer, C.H., Bramble, M.S., Mustard, J.F., 2018. Origin and Emplacement of the Circum-Isidis Olivine-Rich Unit. Presented at the LPSC XXXXIX, LPI, Houston, TX, p. Abstract #1545.
Langevin, Y., Poulet, F., Bibring, J.-P., Schmitt, B., Doute, S., Gondet, B., 2005. Summer Evolution of the North Polar Cap of Mars as Observed by OMEGA/Mars Express. Science 307, 1581–1584.
Lanza, N.L., Fischer, W.W., Wiens, R.C., Grotzinger, J., Ollila, A.M., Cousin, A., Anderson, R.B., Clark, B.C., Gellert, R., Mangold, N., Maurice, S., Le Mouélic, S., Nachon, M., Schmidt, M., Berger, J., Clegg, S.M., Forni, O., Hardgrove, C., Melikechi, N., Newsom, H.E., Sautter, V., 2014. High manganese concentrations in rocks at Gale crater, Mars. Geophys. Res. Lett. 2014GL060329. https://doi.org/10.1002/2014GL060329

Marion, G.M., Crowley, J.K., Thomson, B.J., Kargel, J.S., Bridges, N.T., Hook, S.J., Baldridge, A., Brown, A.J., Ribeiro da Luz, B., de Souza Filho, C.R., 2009. Modeling aluminum-silicon chemistries and application to Australian acidic playa lakes as analogues for Mars. Geochim. Cosmochim. Acta 73, 3493–3511.
McCauley, J.F., Carr, M., Cutts, J.A., Hartmann, W.K., Masursky, H., Milton, D.J., Sharp, R.P., Wilhelms, D.E., 1972. Preliminary mariner 9 report on the geology of Mars. Icarus 17, 289–327.
McSween, H.Y., Labotka, T.C., Viviano-Beck, C.E., 2014. Metamorphism in the Martian crust. Meteorit. Planet. Sci. 50, 1–14. https://doi.org/10.1111/maps.12330
McSween, H.Y., Wyatt, M.B., Gellert, R., Bell III, J.F., Morris, R.V., Herkenhoff, K.E., Crumpler, L.S., Milam, K.A., Stockstill-Cahill, K.R., Tornabene, L.L., Arvidson, R.E., Bartlett, P., Blaney, D., Cabrol, N.A., Christensen, P.R., Clark, B.C., Crisp, J.A., Des Marais, D.J., Economou, T., Farmer, J.D., Farrand, W.H., Ghosh, A., Golombek, M., Gorevan, S., Greeley, R., Hamilton, V.E., Johnson, J.R., Joliff, B.L., Klingelhoefer, G., Knudson, A.T., McLennan, S.M., Ming, D.W., Moersch, J.E., Rieder, R., Ruff, S.W., Schroder, C., de Souza, P.A., Squyres, S.W., Wanke, H., Yen, A.S., Zipfel, J., 2006. Characterization and petrologic interpretation of olivine−rich basalts at Gusev Crater, Mars. J. Geophys. Res. 111.
Michalski, J.L., Niles, P.B., 2010. Deep crustal carbonate rocks exposed by meteor impact on Mars. Nat. Geosci.
Moody, J.B., 1976. Serpentinization: a review. Lithos 9, 125–138.
Morris, R.V., Klingelhofer, G., Bernhardt, B., Schroder, C., Rodionov, D.S., de Souza, P.A., Jr., Yen, A., Gellert, R., Evlanov, E.N., Foh, J., Kankeleit, E., Gutlich, P., Ming, D.W., Renz, F., Wdowiak, T., Squyres, S.W., Arvidson, R.E., 2004. Mineralogy at Gusev Crater from the Mossbauer Spectrometer on the Spirit Rover. Science 305, 833–836.
Morris, R.V., Ruff, S.W., Gellert, R., Ming, D.W., Arvidson, R.E., Clark, B.C., Golden, D.C., Siebach, K., Klingelhofer, G., Schroder, C., Fleischer, I., Yen, A.S., Squyres, S.W., 2010. Identification of Carbonate-Rich Outcrops on Mars by the Spirit Rover. Science science.1189667.
Murchie, S., Arvidson, R., Bedini, P., Beisser, K., Bibring, J.-P., Bishop, J., Boldt, J., Cavender, P., Choo, T., Clancy, R.T., Darlington, E.H., Des Marais, D., Espiritu, R., Fort, D., Green, R., Guinness, E., Hayes, J., Hash, C., Heffernan, K., Hemmler, J., Heyler, G., Humm, D., Hutcheson, J., Izenberg, N., Lee, R., Lees, J., Lohr, D., Malaret, E., T., M., McGovern, J.A., McGuire, P., Morris, R., Mustard, J., Pelkey, S., Rhodes, E., Robinson, M., Roush, T., Schaefer, E., Seagrave, G., Seelos, F., Silverglate, P., Slavney, S., Smith, M., Shyong, W.-J., Strohbehn, K., Taylor, H., Thompson, P., Tossman, B., Wirzburger, M., Wolff, M., 2007. Compact Reconnaissance Imaging Spectrometer for Mars (CRISM) on Mars Reconnaissance Orbiter (MRO). J. Geophys. Res. 112, E05S03, doi:10.1029/2006JE002682.
Mustard, J.F., Ehlmann, B.L., Murchie, S.L., Poulet, F., Mangold, N., Head, J.W., Bibring, J.P., Roach, L.H., 2009. Composition, Morphology, and Stratigraphy of Noachian Crust around the Isidis basin. J. Geophys. Res. 114, doi:/10.1029/2009JE003349.
Mustard, J.F., Pieters, C.., 1989. Photometric phase functions of common geologic minerals and applications to quantitative analysis of mineral mixture reflectance spectra. J. Geophys. Res. 94, 13619–13634.
Mustard, J.F., Poulet, F., Gendrin, A., Bibring, J.-P., Langevin, Y., Gondet, B., Mangold, N., Bellucci, G., Altieri, F., 2005. Olivine and Pyroxene Diversity in the Crust of Mars. Science 307, 1594–1597.
Mustard, J.F., Poulet, F., Head, J.W., Mangold, N., Bibring, J.P., Pelkey, S.M., Fassett, C.I., Langevin, Y., Neukum, G., 2007. Mineralogy of the Nili Fossae region with OMEGA/Mars Express data: 1. Ancient impact melt in the Isidis Basin and implications for the transition from the Noachian to Hesperian. J. Geophys. Res. 112.
Nelder, J.A., Mead, R., 1965. A simplex method for function minimization. Comput. J. 7, 308–313.
Niles, P.B., Catling, D.C., Berger, G., Chassefière, E., Ehlmann, B.L., Michalski, J.R., Morris, R., Ruff, S.W., Sutter, B., 2012. Geochemistry of Carbonates on Mars: Implications for Climate History and Nature of Aqueous Environments. Space Sci. Rev. 174, 301–328. https://doi.org/10.1007/s11214-012-9940-y
Niles, P.B., Leshin, L.A., Guan, Y., 2005. Microscale carbon isotope variability in ALH84001 carbonates and a discussion of possible formation environments. Geochim. Cosmochim. Acta 69, 2931–2944. https://doi.org/10.1016/j.gca.2004.12.012
Ody, A., Poulet, F., Bibring, J.P., Loizeau, D., Carter, J., Gondet, B., Langevin, Y., 2013. Global investigation of olivine on Mars: Insights into crust and mantle compositions. J. Geophys. Res. Planets 118, 234–262.
Osinski, G.R., Tornabene, L.L., Banerjee, N.R., Cockell, C.S., Flemming, R., Izawa, M.R.M., McCutcheon, J., Parnell, J., Preston, L.J., Pickersgill, A.E., Pontefract, A., Sapers, H.M., Southam, G., 2013. Impact-generated hydrothermal systems on Earth and Mars. Icarus, Terrestrial Analogs for Mars: Mars Science Laboratory and Beyond 224, 347–363. https://doi.org/10.1016/j.icarus.2012.08.030

Paukert, A.N., Matter, J.M., Kelemen, P.B., Shock, E.L., Havig, J.R., 2012. Reaction path modeling of enhanced in situ CO2 mineralization for carbon sequestration in the peridotite of the Samail Ophiolite, Sultanate of Oman. Chem. Geol. 330–331, 86–100. https://doi.org/10.1016/j.chemgeo.2012.08.013

Pollack, J.B., Kasting, J.F., Richardson, S.M., Poliakoff, K., 1987. The case for a wet, warm climate on early Mars. Icarus 71, 203–224.

Poulet, F., Gomez, C., Bibring, J.-P., Langevin, Y., Gondet, B., Pinet, P., Belluci, G., Mustard, J., 2007. Martian surface mineralogy from Observatoire pour la Minéralogie, l'Eau, les Glaces et l'Activité on board the Mars Express spacecraft (OMEGA/MEx): Global mineral maps. J. Geophys. Res. 112, 10.1029/2006JE002840.

Ritzer, J.A., Hauck, S.A., 2009. Lithospheric Structure and Tectonics at Isidis Planitia, Mars. Icarus 201, 528–539.

Rogers, D., Warner Nicholas H., Golombek Matthew P., Head James W., Cowart Justin C., 2018. Areally Extensive Surface Bedrock Exposures on Mars: Many Are Clastic Rocks, Not Lavas. Geophys. Res. Lett. 45, 1767–1777. https://doi.org/10.1002/2018GL077030

Ruff, S.W., Christensen, P.R., 2002. Bright and dark regions on Mars: Particle size and mineralogical characteristics based on Thermal Emission Spectrometer data. J. Geophys. Res.-Planets 107, art. no.-5127.

Ruff, S.W., Niles, P.B., Alfano, F., Clarke, A.B., 2014. Evidence for a Noachian-aged ephemeral lake in Gusev crater, Mars. Geology G35508.1. https://doi.org/10.1130/G35508.1

Russell, M.J., Barge, L.M., Bhartia, R., Bocanegra, D., Bracher, P.J., Branscomb, E., Kidd, R., McGlynn, S., Meier, D.H., Nitschke, W., Shibuya, T., Vance, S., White, L., Kanik, I., 2014. The Drive to Life on Wet and Icy Worlds. Astrobiology 140403073038009. https://doi.org/10.1089/ast.2013.1110

Salvatore, M.R., Goudge, T.A., Bramble, M.S., Edwards, C.S., Bandfield, J.L., Amador, E.S., Mustard, J.F., Christensen, P.R., 2018. Bulk mineralogy of the NE Syrtis and Jezero crater regions of Mars derived through thermal infrared spectral analyses. Icarus 301, 76–96. https://doi.org/10.1016/j.icarus.2017.09.019

Salvatore, M.R., Mustard, J.F., Head, J.W., Cooper, R.F., Marchant, D.R., Wyatt, M.B., 2013. Development of Alteration Rinds by Oxidative Weathering Processes in Beacon Valley, Antarctica, and Implications for Mars. Geochim. Cosmochim. Acta.

Schon, S.C., Head, J.W., Fassett, C.I., 2012. An overfilled lacustrine system and progradational delta in Jezero crater, Mars: Implications for Noachian climate. Planet. Space Sci. 67, 28–45. https://doi.org/10.1016/j.pss.2012.02.003

Schulte, M., Blake, D., Hoehler, T., McCollom, T., 2006. Serpentinization and its implications for life on the early Earth and Mars. Astrobiology 6, 1531–1074.

Schultz, R.A., Frey, F.A., 1990. A new survey of large multi-ringed impact basins on Mars. J. Geophys. Res. 95, 14175–14189.

Seelos, F.P., Viviano-Beck, C.E., Morgan, M.F., Romeo, G., Aiello, J.J., Murchie, S.L., 2016. CRISM hyperspectral targeted observation PDS product sets - TERs and MTRDRs. Presented at the LPSC 47, Houston, TX, p. Abstract #1783.

Shahrzad, S., Kinch, K.M., Goudge, T.A., Fassett, C.I., Needham, D.H., Quantin‐Nataf, C., Knudsen, C.P., 2019. Crater statistics on the dark-toned, mafic floor unit in Jezero Crater, Mars. Geophys. Res. Lett. 0. https://doi.org/10.1029/2018GL081402

Shkuratov, Y., Starukhina, L., Hoffmann, H., Arnold, G., 1999. A Model of Spectral Albedo of Particulate Surfaces: Implications for Optical Properties of the Moon. Icarus 137, 235–246. https://doi.org/10.1006/icar.1998.6035

Singer, R.B., Roush, T.L., 1985. Effects of temperature on remotely sensed mineral absorption features. J. Geophys. Res. 90, 12434–12444.

Skok, J.R., Mustard, J.F., Tornabene, L.L., Pan, C., Rogers, D., Murchie, S.L., 2012. A spectroscopic analysis of Martian crater central peaks: Formation of the ancient crust. J Geophys Res 117, E00J18.

Stopar, J.D., Jeffrey Taylor, G., Hamilton, V.E., Browning, L., 2006. Kinetic model of olivine dissolution and extent of aqueous alteration on mars. Geochim. Cosmochim. Acta, A Special Issue Dedicated to Larry A. Haskin 70, 6136–6152. https://doi.org/10.1016/j.gca.2006.07.039

Sunshine, J.M., Pieters, C.M., 1998. Determining the composition of olivine from reflectance spectroscopy. J. Geophys. Res.-Planets 103, 13675–13688.

Taylor, S.R., McLennan, S., 2009. Planetary crusts: their composition, origin and evolution. Cambridge University Press.

Tornabene, L.L., Moersch, J.E., McSween, H.Y., Hamilton, V.E., Piatek, J.L., Christensen, P.R., 2008. Surface and crater-exposed lithologic units of the Isidis Basin as mapped by coanalysis of THEMIS and TES derived data products. J. Geophys. Res. Planets 113, E10001. https://doi.org/10.1029/2007JE002988

Trang, D., Lucey, P.G., Gillis-Davis, J.J., Cahill, J.T.S., Klima, R.L., Isaacson, P.J., 2013. Near-infrared optical constants of naturally occurring olivine and synthetic pyroxene as a function of mineral composition. J. Geophys. Res. Planets 118, 708–732.

Usui, T., McSween Harry Y., Clark Benton C., 2008. Petrogenesis of high−phosphorous Wishstone Class rocks in Gusev Crater, Mars. J. Geophys. Res. Planets 113. https://doi.org/10.1029/2008JE003225

van Berk, W., Fu, Y., 2011. Reproducing hydrogeochemical conditions triggering the formation of carbonate and phyllosilicate alteration mineral assemblages on Mars (Nili Fossae region). J. Geophys. Res. Planets 116, E10006. https://doi.org/10.1029/2011JE003886

Van Kranendonk, M.J., Philippot, P., Lepot, K., Bodorkos, S., Pirajno, F., 2008. Geological setting of Earth's oldest fossils in the ca. 3.5†Ga Dresser Formation, Pilbara Craton, Western Australia. Precambrian Res. 167, 93–124.

Viviano, C.E., Moersch, J.E., McSween, H.Y., 2013. Implications for early hydrothermal environments on Mars through the spectral evidence for carbonation and chloritization reactions in the Nili Fossae region. J. Geophys. Res. Planets 118, 1858–1872.

Viviano-Beck, C.E., Seelos, F.P., Murchie, S.L., Kahn, E.G., Seelos, K.D., Taylor, H.W., Taylor, K., Ehlmann, B.L., Wisemann, S.M., Mustard, J.F., Morgan, M.F., 2014. Revised CRISM Spectral Parameters and Summary Products Based on the Currently Detected Mineral Diversity on Mars. J. Geophys. Res. Planets 2014JE004627. https://doi.org/10.1002/2014JE004627

Walter, M.R., Des Marais, D.J., 1993. Preservation of biological information in thermal spring deposits: developing a strategy for the search for fossil life on Mars. Icarus 101, 129–143.

Wänke, H., 1981. Constitution of terrestrial planets. Phil Trans R Soc Lond A 303, 287–302. https://doi.org/10.1098/rsta.1981.0203

Wilhelms, D.E., 1973. Comparison of Martian and lunar multiringed circular basins. J. Geophys. Res. 78, 4084–4095. https://doi.org/10.1029/JB078i020p04084

Willcox, A., Buisman, I., Sparks, R.S.J., Brown, R.J., Manya, S., Schumacher, J.C., Tuffen, H., 2015. Petrology, geochemistry and low-temperature alteration of lavas and pyroclastic rocks of the kimberlitic Igwisi Hills volcanoes, Tanzania. Chem. Geol. 405, 82–101. https://doi.org/10.1016/j.chemgeo.2015.04.012

Williford, K.H., 2018. The Mars2020 Rover - a progress report, in: Mars: From Habitability to Life.

Wray, J.J., Murchie, S.L., Bishop, J.L., Ehlmann, B.L., Milliken, R.E., Wilhelm, M.B., Seelos, K.D., Chojnacki, M., 2016. Orbital evidence for more widespread carbonate-bearing rocks on Mars. J. Geophys. Res. Planets 2015JE004972. https://doi.org/10.1002/2015JE004972